\documentclass[conference]{IEEEtran}
\IEEEoverridecommandlockouts
\usepackage{cite}
\usepackage{amsmath,amssymb,amsfonts}
\newtheorem{defn}{Definition}
\usepackage{algorithmic}
\usepackage{graphicx}

\usepackage{textcomp}
\usepackage{threeparttable}
 \usepackage{multirow}
 \usepackage{diagbox}
 \usepackage{booktabs}
\usepackage[ruled,linesnumbered]{algorithm2e}
\usepackage{bbm}
\usepackage{subfigure}

\usepackage{xcolor}
\def\BibTeX{{\rm B\kern-.05em{\sc i\kern-.025em b}\kern-.08em
    T\kern-.1667em\lower.7ex\hbox{E}\kern-.125emX}}
\begin{document}

\title{CADRL: Category-aware Dual-agent Reinforcement Learning for Explainable Recommendations over Knowledge Graphs \\
}
\author
{
	Shangfei Zheng{\small$^\dag$}\hspace*{10pt}
  	
	Hongzhi Yin{\small$^\ddag$}{$^*$}\hspace*{10pt}\thanks{* Corresponding Author.}
	Tong Chen{\small$^\ddag$}\hspace*{10pt}
 
 Xiangjie Kong {\small$^\ddag\ddag$} \hspace*{10pt}
	Jian Hou{\small$^\dag $}\hspace*{10pt}\hspace*{10pt}   Pengpeng Zhao {\small$^\dag\dag$} \hspace*{10pt}\\
	\fontsize{10}{10}\selectfont\itshape $~^\dag$School of Computer Science and Technology, Zhejiang Sci-Tech University\\

	\fontsize{10}{10}\selectfont\itshape
	$~^\ddag$ School of Information Technology and Electrical Engineering, The University of Queensland\\
 \fontsize{10}{10}\selectfont\itshape

 $~^\ddag\ddag$ College of Computer Science and Technology, Zhejiang University of Technology\\

	$~^\dag\dag$ School of Computer Science and Technology, Soochow University\\
		\fontsize{10}{10}\selectfont\itshape

	\fontsize{9}{9}\selectfont\ttfamily\upshape$~^\dag$zsf@zstu.edu.cn\hspace*{10pt}$~^\ddag$db.hongzhi@gmail.com\hspace*{10pt}$~^\ddag$tong.chen@uq.edu.au\\ 
 \hspace*{10pt}$~^\ddag\ddag$xjkong@ieee.org
 \hspace*{10pt}
$~^\dag$changeleap@163.com \hspace*{10pt}$~^\dag\dag$ppzhao@suda.edu.cn\\
}
\maketitle

\begin{abstract}
Knowledge graphs (KGs) have been widely adopted to mitigate data sparsity and address cold-start issues in recommender systems. While existing KGs-based recommendation methods can predict user preferences and demands, they fall short in generating explicit recommendation paths and lack explainability. As a step beyond the above methods, recent advancements utilize reinforcement learning (RL) to find suitable items for a given user via explainable recommendation paths. However, the performance of these solutions is still limited by the following two points. (1) Lack of ability to capture contextual dependencies from neighboring information. (2) The excessive reliance on short recommendation paths due to efficiency concerns. To surmount these challenges, we propose a category-aware dual-agent reinforcement learning (CADRL) model for explainable recommendations over KGs. 
Specifically, our model comprises two components: (1) a category-aware gated graph neural network that jointly captures context-aware item representations from neighboring entities and categories, and (2) a dual-agent RL framework where two agents efficiently  traverse long paths to search for suitable items. Finally, experimental results show that CADRL outperforms state-of-the-art models in terms of both effectiveness and efficiency on large-scale datasets.

\end{abstract}


\section{Introduction}
Recommender systems have emerged as indispensable tools for personalized information retrieval by providing users with suitable items \cite{survey2} \cite{survey3} \cite{EX5}. Despite their widespread adoption in the industry, recommender systems suffer from the limitations of data sparsity and cold starts \cite{Ripplenet}, e.g., some items are purchased or viewed by only a few users or none at all. In this regard, researchers have equipped recommender systems with the capability to leverage knowledge graphs (KGs), aiming to effectively integrate diverse structural knowledge to alleviate the limitations \cite{KGAT}. The uniqueness of KG-based recommender systems lies in mapping users, items, and attributes (e.g., brand or feature) as entities and utilizing semantic relations (e.g., also\_viewed or purchased) to link various types of entities \cite{MetaKG}. By integrating these structural data into the recommendation process, KG-based recommendation methods not only improve the recommendation accuracy but also establish a foundation for the traceability of results \cite{EMKR}.

In the literature, there has been a long line of studies in KG-based recommendation methods, such as embedding-based methods \cite{embedding1} \cite{embedding2} and neural network-based techniques \cite{NN1} \cite{NN2}. While these recommendation methods leverage structural information to achieve notable progress, they act as black-box systems that lack explainability \cite{PGPR} \cite{ADAC}. To enhance the explainability, path-based recommendation methods \cite{HeteroEmbed} \cite{KGRSrules} first traverse entities and relations over KGs and then capture the connection patterns in user-item pairs, which can explicitly model user preferences for items \cite{path1} \cite{path2} \cite{SAPL}. 
However, these traditional path-based methods cannot explore all paths for each user-item pair due to the lack of dynamic decision-making ability \cite{INFER}. 

Recent advancements in KG-based recommender systems reveal that reinforcement learning (RL)-based recommendation methods achieve promising and explainable results over KGs \cite{ReMR}. Unlike traditional path-based methods, existing RL-based methods transform the recommendation task into a multi-hop reasoning process where the target is to train an agent to make optimal decisions (i.e., optimal selection of paths) and maximize the accumulated rewards (e.g., finding suitable items). Representatively, PGPR\cite{PGPR} trains an agent guided by a multi-hop scoring function to derive dynamic decisions from a given user to appropriate items. ADAC \cite{ADAC} builds upon the technical framework of PGPR, further integrating the generative adversarial imitation learning to enhance its convergence rate. Furthermore, subsequent work primarily focuses on carefully designed RL frameworks \cite{UCPR} \cite{INFER} and convincing explanations \cite{CogER} \cite{ReMR}. However, the performance of these methods is still limited by the following two challenges.

\begin{figure}
 \centering 
  \label{Fig2}
  \includegraphics[width=0.95\linewidth,height=6cm]{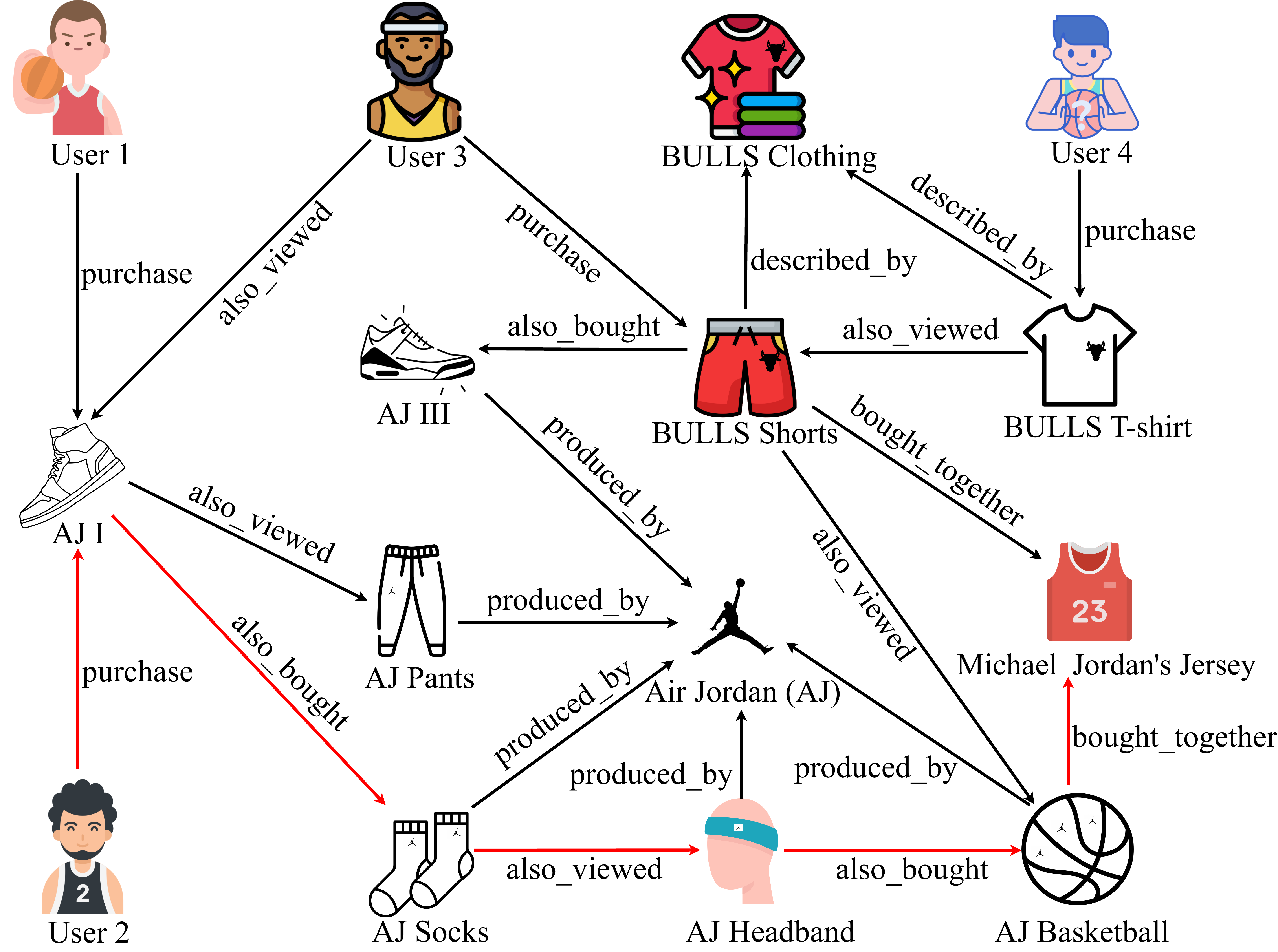}
 \hfill 
 \caption{
A small fragment of a knowledge graph formed by the interactions of multiple users with the same shopping preferences. The red arrow is a 5-hop recommendation path starting from User 2 to a recommended item. To easily follow our work, we exemplify this fragment throughout the paper. 
}
\vspace{-0.4cm}
\end{figure}


\textbf{Challenge I}. The inability to capture contextual dependencies from neighboring information reduces the effectiveness of the item representation in the aforementioned RL-based methods.
In fact, items with supplementary descriptions emerge as the most important entity type in KG-based recommender systems by virtue of their direct impact on recommendation results \cite{survey2}. It is essential to capture valuable representations of items during the recommendation process. Specifically, (1) contextual dependencies from neighboring entities provide high-order semantic information to the current item's representations. On the one hand, the contextual dependencies from neighboring entities of the item type can capture the connection patterns and contextual semantics. By way of illustration, the item ``BULLS Shorts'' enhances the potential relevance of basketball and sports in the semantics by receiving the contextual information from the neighboring entities of the item type ``AJ III''  and ``Michael Jordan's Jersey'' in Fig.1. On the other hand, neighboring entities of the attribute type provide more supplementary information for the semantics of the current item. As shown in Fig. 1,  the connection between ``BULLS Shorts'' and the attribute entity ``BULLS Clothing'' indicates that the Shorts are a piece of clothing associated with BULLS. (2) Contextual dependencies from neighboring item-categories contain the inherent meta-data of neighboring items, reflecting the transfer pattern of neighboring items. For example, ``BULLS Shorts'' are closely linked to the neighboring category of basketball equipment in the KG, which highlights cross-selling opportunities or complementary relationships. Obviously, capturing contextual dependencies from neighboring entities and item-categories simultaneously enhance the effectiveness of item representation.

\textbf{Challenge II}.
The second challenge is that the existing RL-based  methods rely heavily on short recommendation chains (i.e., the maximum length of a recommendation path is commonly limited to 3) due to efficiency concerns. This limitation undermines the effectiveness of the recommendation policies and fails to satisfy the demand for long-sequence recommendations \cite{path1}. Specifically, (1) short path lengths can slightly alleviate the issue of inefficient recommendations by reducing the combination of semantic relations over KGs \cite{PGPR}, but they limit the optimization of recommendation policies. First, the stability of the recommendation policy diminishes when these methods are unable to infer potential items beyond three-hop paths \cite{MMKGR}. For instance, the target item ``Michael Jordan's Jersey'' is inferred via a 5-hop path, while the 3-hop item ``AJ Headband'' is not the answer for ``User 2''.  Second, short paths involving only a few entities and relations are inadequate to precisely capture users' interests and preferences, which leads to false results \cite{multihopKGR} and misleads recommendation policies \cite{KGAT}. (2) In reality, users frequently browse and purchase products across multiple item-categories \cite{CoCoRec}. This practical demand necessitates recommender systems to adopt long recommendation paths to   capture users' evolving interests across categories. However, it is non-trivial to directly extend the above single-agent RL-based recommendation methods \cite{PGPR} \cite{ADAC} \cite{UCPR}  \cite{INFER} \cite{SAPL} \cite{ReMR} \cite{CogER} to  scenarios involving recommendation long paths. This is because they tend to generate large action spaces as the path lengths increase, thereby impairing recommendation efficiency \cite{multiagentadvan}. To make matters worse, these methods struggle to rapidly acquire sufficient reward signals over long recommendation paths, exacerbating the negative impact of the sparse reward dilemma on the RL framework \cite{RLsurvey1}. Consequently, it is crucial to efficiently use long paths and improve the accuracy of explainable recommendations.


In light of the aforementioned challenges, we propose a novel model entitled CADRL  (\textbf{C}ategory-\textbf{a}ware \textbf{D}ual-Agent
\textbf{R}einforcement \textbf{L}earning). It draws inspiration from two key insights: (1) high-order entity representations can be derived from neighboring entities and categories in KGs \cite{KGsurvey}; (2) multi-agent RL with collaborative decision-making capabilities excels at efficiently finding targets over long paths \cite{MARLsurvey} \cite{EX7}. A notable distinction of our model from existing KG-based recommendation methods is that CADRL elegantly captures high-order representations of items from neighboring information as well as conducts explainable recommendations via a novel dual-agent RL framework. Specifically, CADRL contains two components. (1) To solve \textbf{Challenge I}, a category-aware gated graph neural network (CGGNN) is designed to generate context-aware  item representations by jointly capturing contextual dependencies from neighboring entities and item-categories. Its gated graph neural network module adeptly extracts semantic information with low noise from the entity level. Then, a category-aware graph attention network module obtains shared features from the neighboring item-category level to form the final item representation. (2) To solve \textbf{Challenge II}, we propose a dual-agent reinforcement learning (DARL) framework to conduct explainable recommendations and efficiently discover suitable items through the collaborative reward mechanism and shared policy networks. In conclusion, this paper presents the following contributions:

\begin{itemize}
\item To the best of our knowledge, we are the first to point out the necessity of  exploiting contextual dependencies from neighboring information and the need for designing dual-agent RL to efficiently conduct explainable recommendations without relying solely on short paths.
\item  We propose CADRL which consists of components named CGGNN and DARL. The former generates high-order context-aware representations for items, while the latter efficiently performs explainable recommendations via long paths over KGs.
\item We conduct extensive experiments on several real-world benchmark datasets. The results showcase the superiority of CADRL compared to state-of-the-art baselines.
\end{itemize}

\section{Related Work}


\subsection{Reinforcement Learning}

Reinforcement learning is a machine learning paradigm in which one or more agents learn to make decisions by interacting with the environment using MDPs \cite{RLsurvey1}. Typically, the goal of RL is to use a policy network to approximate the decisions taken in different states to maximize the long-term cumulative reward \cite{RLsurvey2}. Existing RL methods can be divided into two categories: single-agent RL and multi-agent RL methods \cite{MARLsurvey1}. In the former, only a single agent interacts with the environment during the learning process, and its decision-making is hindered by the limitations of the action space and sparse rewards \cite{MARLsurvey2}. The latter trains collaborative decision-making abilities and shares collective intelligence for multiple agents to address  the issue of large action spaces \cite{MARLsurvey3}, thereby efficiently using long sequences to complete tasks \cite{MARLsurvey}.


\subsection{KG-based Recommendation}
A KG is a semantic network composed of entities and relations, it has been widely used in recommender systems \cite{KGzsf}. KG-based recommender systems, designed to fulfill the personalized demands of users, map the connections between users and items into a semantic structure \cite{inf}. Most existing KG-based recommendation methods first transform user-item and item-item pair interactions into semantic relations between entities in KGs \cite{KGRS1} \cite{EX1} \cite{EX3} \cite{EX4}, and then fully exploit the implicit patterns in structural information to provide more accurate recommendation results for users. Generally, existing KG-based recommendation methods can be broadly categorized into two types: embedding-based and path-based methods \cite{KGRSsurveyrw}. Specifically, embedding-based methods  utilize knowledge graph embedding \cite{KGEsurvey} or neural network techniques\cite{gnnsurvey} \cite{wei} \cite{EX2} to enhance the representation of items and users. These methods demonstrate that KG-based recommender systems utilizing structured data can  improve recommendation accuracy \cite{NN1}. However, they cannot effectively model the connection patterns  and interpret the recommended results using semantics in KGs \cite{ADAC}. Path-based methods use predefined connection patterns (e.g., meta-paths \cite{metapath}, meta-graphs \cite{metagraph}, and rules \cite{KGRSrules}) to model the paths between users and items for recommendations. These methods capture complex relations among entities to improve the explainability of KG-based recommendation methods but face two limitations: (1) lack of dynamic decision-making capabilities; (2) fully exploring all paths for each user-item pair is impractical in large-scale KGs \cite{PGPR}.

\subsection{RL-based Explainable Recommendations  over KGs}

Recently, RL has been applied to KG-based recommender systems to integrate recommendation and explainability by offering recommendation paths. Related methods transform the recommendation problem of finding suitable items for a user into a multi-hop reasoning problem, starting from an initial user entity to target item entities. Technically, these methods utilize Markov Decision Process (MDP) to dynamically discover the optimal paths over KGs. This decision process is as intuitive as ``taking a walk" over KGs, which naturally yields explainable results for recommender systems. Specifically, PGPR \cite{PGPR} is the pioneering work  applying RL to conduct explainable recommendations over KGs.  It trains a single agent with an innovative soft reward strategy and action pruning strategy to conduct  recommendations and path-finding simultaneously. Subsequent studies build upon this technical idea. ADAC \cite{ADAC} jointly models both demonstrations and historical user preferences to obtain accurate recommendation results with fast convergence. To alleviate the problem of huge action space, UCPR \cite{UCPR}  adopts a multi-view reinforcement learning framework to simplify the path search process from the perspective of user demand. Considering that the entity representation cannot be updated by knowledge reasoning, INFER \cite{INFER} first constructs a subgraph based on the reasoning paths to input the path information into the GNN layer, and then designs a joint learning framework to allow representation and recommendation  modules to enhance each other. Additionally, to address the challenge that existing work ignores the sentiment on relations (e.g., user  satisfaction),  SAPL \cite{SAPL} designs a sentiment-aware reinforcement learning method to conduct accurate and convincing recommendations in sentiment-aware KGs. To further improve the accuracy and the rationality of the recommendation results, ReMR \cite{ReMR} and CogER \cite{CogER} fully explore the structural information of KGs using multi-view RL and multi-system RL, respectively. Although these methods effectively perform explainable recommendations over KGs, they ignore the following details.  (1) Contextual  dependencies from both the entity and category levels can add high-order semantic information to the representation of items.  (2) Single-agent RL methods tend to rely on short paths and ignore the diverse long path combinations, thereby limiting the performance of the recommendations.

\begin{figure*}
 \centering 
 \begin{center}
  \label{Fig2}
  \includegraphics[width=0.99\linewidth,height=5.7cm]{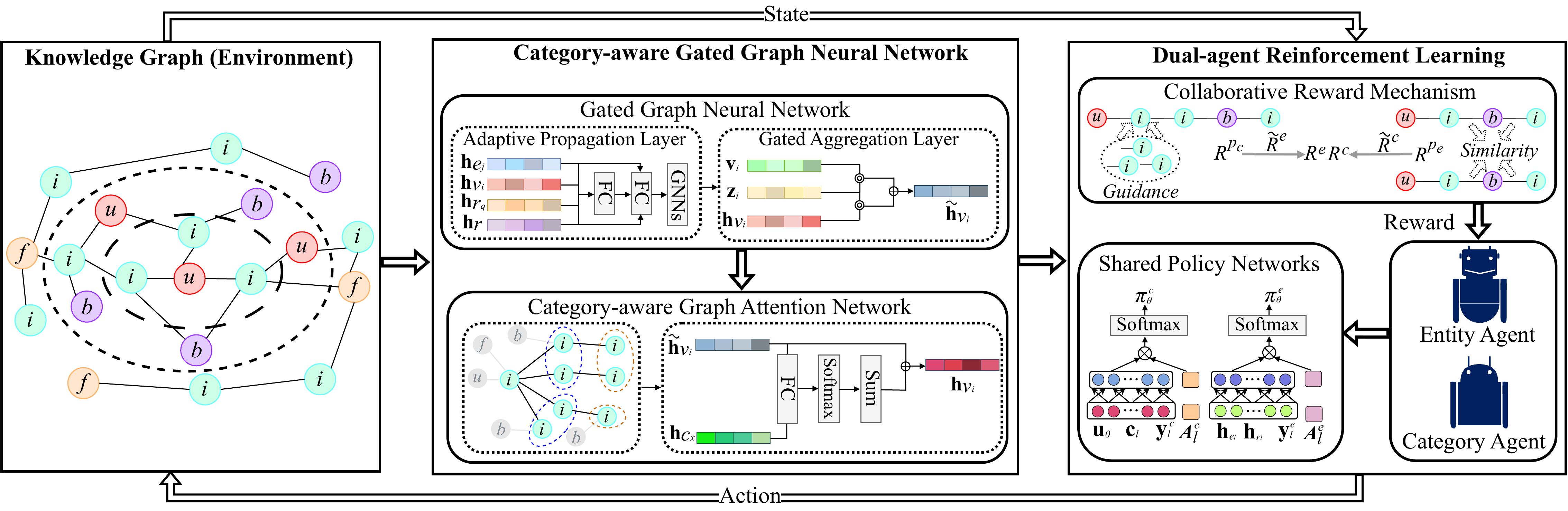}
 \hfill 
 \caption{
Overview of CADRL. Different colored circles represent  various types of entities, including items, brands, features, and users. After obtaining high-order representations of items from CGGNN, DARL fully interacts with the KG. Dual agents receive the state and reward as well as output the next action until the appropriate items are inferred.}
\end{center}
\vspace{-15pt}
\end{figure*}

\section{Preliminaries and Definitions}
A KG  
	\begin{math}
		\mathcal{G} = \{\mathcal{E}, \mathcal{R}, \mathcal{T}\} 
	\end{math} is essentially a multi-relation graph, where $\mathcal{E}$ represents the entity set and $\mathcal{R}$ denotes a range of semantic relations.
	\begin{math}
		\mathcal{T} = \{ (\emph{$e_s$}, \emph{r}, \emph{$e_d$}) \mid \emph{$e_s$}, \emph{$e_d$} \in \mathcal{E}, \emph{r} \in \mathcal{R} \} 
	\end{math} represents a collection of triplets in $\mathcal{G}$, where $e_s$, $e_d$, and $r$ denote the  head entity, tail entity, and the semantic relation between these entities, respectively.  Users, items, and attributes are mapped as entities as well as the interactions connecting them are modeled as semantic relations in the scenario of KG-based recommendation. Following previous studies \cite{UCPR} \cite{CAFE}, the entity set $\mathcal{E}$ includes subsets such as the user set $\mathcal{U}$, the item set $\mathcal{V}$, the feature set $\mathcal{F}$, and the brand set $\mathcal{B}$, where $\mathcal{U}$, $\mathcal{V}$, $\mathcal{F}$, $\mathcal{B}$ $\subseteq$ $\mathcal{E}$, $\mathcal{U}$  $\cap$  $\mathcal{V}$ $\cap$  $\mathcal{F}$ $\cap$  $\mathcal{B}$ =  $\varnothing$. The relation set $\mathcal{R}$ comprises  various types of relations (e.g.  purchase or also\_bought) that capture semantic connections between entities.  Following previous work \cite{CogER}, the inverse relation is also appended to the triplets, i.e., each triplet (\emph{$e_s$}, \emph{$r$}, \emph{$e_d$}) is equivalent to the triplet (\emph{$e_d$}, \emph{$r^{-1}$}, \emph{$e_s$}). Without loss of generality, we ensure the reachability of recommendation paths by converting (?, \emph{$r$}, \emph{$e_d$}) to (\emph{$e_d$}, \emph{$r^{-1}$}, ?). To facilitate a deep understanding of our methodology, we present our definitions as follows.

 \begin{defn}
	\label{def:activity}
		\emph{Multi-hop Reasoning}. Given a query $($\emph{$e_s$}, \emph{$r$}, ?$)$, where ``?" denotes the missing entity in KGs, the goal of multi-hop reasoning is to infer the missing entity by relation paths shorter or equal $L$ hops, where $L \in \mathbb{Z}_{\geq 1}$, $\mathbb{Z}$ represents the set of integers.

\end{defn}

 \begin{defn}
 \emph{Contextual Dependency in KGs}.  Contextual dependency at the entity and entity-category levels reveals the inherent influence of neighboring topology and high-order semantics on entity representation in KGs. 
 Given an entity $v_i$ and its representation $\textbf{h}_{v_i}$, the context-dependent processes propagated by its neighboring entity set $\mathcal{N}_{v_i}$ and neighboring category set $\mathcal{N}^c_{v_i}$ are  expressed respectively as $\textbf{h}_{v_i}$ $\leftarrow$ $f_e$($\mathcal{N}_{v_i}$), $\textbf{h}_{\emph{v}_{\emph{i}}}$ $\leftarrow$ $f_c$($\mathcal{N}^c_{v_i}$), where $f_e$ and $f_c$ are propagation functions.

\end{defn}

\begin{defn}
 \emph{Markov Decision Process}. MDP is a mathematical framework in RL used to model  the interaction between agents and their environment. It includes the state space $S$, action space $A$,  transition function $P$, and reward function $R$. The agent selects an action $a \in A$, transitions to the next state according to $P$, and receives a reward value.
\end{defn}

\begin{defn}
 \emph{Category Knowledge Graph}. The category knowledge graph $\mathcal{G}^c$ can be viewed as a dense virtual mapping of KG $\mathcal{G}$. It is represented as a directed graph $\mathcal{G}^c = (\mathcal{V}^c, \mathcal{R}, \mathcal{T}^c)$, where $\mathcal{V}^c$ represents categories of items in $\mathcal{G}$, $\mathcal{R}$ denotes semantic relations, and $\mathcal{T}^c$ represents a collection of triplets in $\mathcal{G}^c$. Two categories are connected in $\mathcal{G}^c$ if there is at least one relation $r$ $\in$ $\mathcal{R}$ between entities in the two categories. $\mathcal{G}^c$ captures hierarchical and taxonomic category association, providing high-level  semantic knowledge. 
\end{defn}

\textbf{Problem}: Our work focuses on RL-based explainable recommendations over KGs and uses RL technology to infer the recommended item set $\mathcal{V}_u$ = \{$v_{u,i}|i=1,...,n$\} for a specific user $u$ $\in$ $\mathcal{U}$ through multi-hop reasoning. Semantic combinations in $L$-hop recommendation paths ``$u$ $\stackrel{r_1}{\longrightarrow}$ $e_1$ $\stackrel{r_2}{\longrightarrow}$ ...  $\stackrel{r_{L}}{\longrightarrow}$ $v_{u,i}$''  provide explainable evidence for why a particular recommendation process is considered. The problem formulation is defined as follows:
	\begin{itemize}
		\item Our inputs comprise the user set $\mathcal{U}$, the item set $\mathcal{V}$, the observed interactions $\mathcal{V}_u$, and  $\mathcal{G}$.
		\item Our outputs are the recommended item set $\mathcal{V}_u$ = \{$v_{u,i}|i=1,...,n$\} and  corresponding $L$-hop recommendation paths $\tau$($u$, $v_{u,i}$).
	\end{itemize}

\section{Methodology}
\subsection{Overview of CADRL}
As shown in Fig. 2, CADRL contains two components: (1) a category-aware gated graph neural network (CGGNN) that generates high-order item representations by capturing contextual dependencies from neighboring entities and categories in a low-noise manner; (2) a dual-agent reinforcement learning (DARL), which leverages collaborative agents to efficiently conduct explainable  recommendation based on the high-order representation  obtained from CGGNN. In what follows, we elaborate on the design of both components.

\subsection{Category-aware Gated Graph Neural Network}
Existing methods  \cite{PGPR} \cite{ADAC} \cite{UCPR}  \cite{INFER} \cite{SAPL} \cite{ReMR} \cite{CogER} are powerless to capture contextual dependencies from neighboring entities and categories simultaneously, which limits the effectiveness of item representation \cite{DREAM} \cite{TNNLS}. To address the problem, we introduce CGGNN in this subsection. It jointly learns the semantics and topology of neighboring items and categories, generating high-order item representations in a low-noise manner. Compared with existing representation methods in the field of KG-based recommendation, the main technical difference of CADRL lies in the following two aspects. (1) We comprehensively consider the contextual dependencies from neighboring entities and categories, thereby providing extensive semantic and topological information for high-order item representations. (2) To ensure the effectiveness of high-order item representation, we incorporate a gated recurrent unit and an attention mechanism to minimize noise interference. Notably, CGGNN exclusively generates high-order  representations for items without extending to other entity types. This design choice is motivated by three main factors. (1) Items directly influence recommendation outcomes
and are pivotal within KG-based recommender systems \cite{ITEM}.  (2) Items have an obvious semantic hierarchy than other types of entities \cite{survey2}. (3) This design principle minimizes  computational costs during the calculation process, particularly in large-scale KG-based recommender systems.

Specifically, CGGNN includes a gated graph neural network (GGNN) module and a category-aware graph attention network (CGAN) module. Initial representations of all entities are obtained using TransE \cite{TransE}.  GGNN leverages the adaptive propagation layer and gated aggregation layer to capture fine-grained contextual dependencies from neighboring entities. Subsequently, CGAN employs an attention mechanism to perform weighted calculations on common features among neighboring categories, thereby generating high-order context-aware representations for items.

\subsubsection{Gated Graph Neural Network}
Typically, existing knowledge representation techniques utilize Graph Neural Networks (GNNs) to capture multi-hop contextual features in the field of knowledge graph completion \cite{KGC2} \cite{KGC1}. However, these solutions are not suitable for RL-based explainable recommendations over KG. This is because they overlook the fact that neighboring relations of an item provide varying semantic strength  for shopping behavior \cite{GNNDIS1}. For instance, the contextual information provided by ``AJ Basketball $\stackrel{bought\_together}{\longrightarrow}$ Michael Jordan's Jersey'' holds more significance than ``AJ Basketball $\stackrel{produced\_by}{\longrightarrow}$ Air Jordan (AJ)'' when executing recommendations for ``User 2'' in Fig.1. To make matters worse, existing GNN-based representation methods introduce noise into the recommendation process due to the semantic decay  among multi-hop neighboring entities \cite{GGNNnoise}. To address these issues, we design GGNN including an adaptive propagation layer and a gated aggregation layer. 


\textbf{Adaptive Propagation Layer:} we first obtain the representation $\textbf{t}_{i,r,j}$  of the triplet ($v_i$, $r$, $e_j$), which encompasses sufficient topological information by concatenating the entity  representations $\textbf{h}_{v_i}$,  $\textbf{h}_{e_j}$  and their relation representation $\textbf{h}_r$\cite{triplet}. Noted here, \emph{$e_j$} belongs to other entity types besides user, i.e.,  \emph{$e_j$} $\in$ $\mathcal{E}$ $\cup$ $\mathcal{V}$ $\cup$ $\mathcal{F}$ $\cup$ $\mathcal{B}$. Then, we additionally integrate the purchase relation $\textbf{h}_{r_p}$ into the triplet representation $\textbf{t}_{i,r,j}$ to  assess the semantic strength of  relevant relations with shopping behavior in context dependency. Based on this, we calculate semantic correlation between $\alpha_{i,r}$. Next, we define the information propagation of neighboring entities in the $k$-th GNN layer. The above process is defined as follows, 
\begin{equation}
 \textbf{t}_{i,r,j} = \sigma (\textbf{W}_1 [ \textbf{h}_{v_i} \oplus \textbf{h}_{e_j} \oplus \textbf{h}_r\oplus \textbf{h}_{r_p}])
\end{equation}
\begin{equation}
 \alpha_{i,r} = \sigma (\textbf{W}_2 \textbf{t}_{i,r,j} + \textbf{b}) 
\end{equation}
\begin{equation}
\begin{split}
 \textbf{n}_{v_i}^{k} = \sum_{(r,j) \in \mathcal{N}_{i}(v_i)} \alpha_{i,r} \textbf{W}_{in}^{(k-1)} (\textbf{h}_{e_j}^{(k-1)} \circ \textbf{h}_r^{(k-1)}) \\ + \sum_{(r,j) \in \mathcal{N}_{o}(v_i)} \alpha_{i,r} \textbf{W}_{out}^{(k-1)}  (\textbf{h}_{e_j}^{(k-1)} \circ \textbf{h}_r^{(k-1)})
 \end{split}
\end{equation} 
where $\oplus$ and $\circ$ represent the concatenation and  element-wise product, respectively. $\sigma$ denotes  the $sigmoid$ activation function. In addition, $\alpha_{i,r}$ is the attention weight of the triplet ($v_i$, $r$, $e_j$).  $\mathcal{N}_{i}$ and $\mathcal{N}_{o}$ represent the incoming and outgoing neighbors of $v_i$, respectively. $\textbf{W}_{in}^{k-1}$ and $\textbf{W}_{out}^{k-1}$ $\in$ $\mathbb{R}^{d \times d}$ denote the transformation matrices, respectively. Based on this, we obtain the representation $\textbf{n}_{v_i}^{k}$ of contextual dependencies passed from neighboring entities to the item $v_i$. 

\textbf{Gated Aggregation Layer:}  After obtaining the above representation,  the gated aggregation layer fuses $\textbf{n}_{v_i}$ with the self-embedding $\textbf{h}_{v_i}$ to update the item's representation.  Furthermore, we employ the Gated Recurrent Units (GRUs) to eliminate the noise introduced by the semantic decay of multi-hop neighbor entities in the update process. The calculation process is as follows:
\begin{equation}
 \textbf{z}_{i}^{(k)} = \sigma ( \textbf{W}_{z1}  \textbf{n}_{v_i}^{(k)} + \textbf{W}_{self}\textbf{h}_{v_i}^{(k-1)})
\end{equation}
\begin{equation}
\hat{\textbf{v}}_{i}^{(k)} = \sigma ( \textbf{W}_{v1}  \textbf{n}_{v_i}^{(k)} + \textbf{W}_{v2}\textbf{h}_{v_i}^{(k-1)} )
\end{equation}
\begin{equation}
 \textbf{v}_{i}^{(k)}  = tanh ( \textbf{W}_{\hat{v}1}  \textbf{n}_{v_i}^{(k)} + \textbf{W}_{\hat{v}2}( \hat{\textbf{v}}_{i}^{(k)}  \circ \textbf{h}_{v_i}^{(k-1)})) 
\end{equation}
\begin{equation}
\Tilde{\textbf{h}}_{v_i}^{(k)} = (\textbf{1}- \textbf{z}_{i}^{(k)})\circ  \textbf{h}_{v_i}^{(k-1)} +  \textbf{z}_{i}^{(k)}\circ \textbf{v}_{i}^{(k)}  
\end{equation}
where $\textbf{W}_{z1}$, $\textbf{W}_{v1}$, $\textbf{W}_{\hat{v}1}$, $\textbf{W}_{self}$, $\textbf{W}_{v2}$, $\textbf{W}_{\hat{v}2}$ $\in$ $\mathbb{R}^{d \times d}$ are learnable parameters. $\textbf{z}_{i}$ and $\hat{\textbf{v}}_{i}$ $\in$ $\mathbb{R}^{d}$ are the update gate and reset gate, dictating the preservation and filtration of information during the fusion process. The item  representation $\Tilde{\textbf{h}}_{v_i}$ that fully captures contextual dependency from neighboring
entities is the trade-off embedding between $\textbf{h}_{v_i}$ and $\textbf{n}_{v_i}$. 

\subsubsection{Category-aware Graph Attention Network}
Existing recommendation methods over KGs often ignore contextual dependencies from category-level, thereby failing to capture common item attributes \cite{category01}. To solve the above problem, we introduce CGAN,  aimed at learning high-order semantic information from neighboring item-categories. Note that, CGAN only calculates the category information for items, as other entity types lack obvious semantic hierarchies in KGs \cite{KGRSsurveyrw}.

Following the existing work \cite{PGPR}, we initialize item-categories $C_1$, $C_2$...$C_n$ using predefined category labels and then employ the mean embedding of all items within each item-category as the category representation. Subsequently, we utilize the $LeakyReLU$ activation function to calculate the aggregation coefficient $\beta_{v_{i}c_{\emph{x}}}$ between the current item $v_i$ and its neighboring  categories in the $m$-th GNN layer:
\begin{equation}
\beta_{v_{i}c_{\emph{x}}}^{(m-1)} = \rm{LeakyRelu}(\textbf{W}_{\emph{ic}}^{(\emph{m}-1)}(\Tilde{\textbf{h}}_{\emph{v}_\emph{i}}^{(\emph{m}-1)} \oplus \textbf{h}_{\emph{c}_\emph{x}}^{(\emph{m}-1)}))
\end{equation}
where $\emph{c}_\emph{x}$ $\in$ $\mathcal{N}_{v_{i}}^c$ represents  the neighboring category for the item $v_i$, 
 and $\textbf{W}_{\emph{ic}}$ is a weight matrix.

Next, we further calculate the attention weight $\alpha_{\emph{v}_{\emph{i}}c_{\emph{x}}}$ as follows:
\begin{equation}
	\alpha_{\emph{v}_{\emph{i}}c_{\emph{x}}}^{(m-1)} = \frac{\rm{exp}(\beta_{\emph{v}_{\emph{i}}c_{\emph{x}}}^{(\emph{m}-1)})}{\sum\nolimits_{c_{\emph{y}} \in  \mathcal{N}^{c}_{\emph{v}_{\emph{i}}}}\rm{exp}(\beta_{\emph{v}_{\emph{i}}c_{\emph{\emph{y}}}}^{(\emph{m}-1)})}
\end{equation}

Afterwards, we can obtain the contextual dependency from  neighboring categories, 
\begin{equation}
	\textbf{h}_{\emph{v}_{\emph{i}}^{c}}^{(m)} = \sum\limits_{c_{\emph{x}} \in  \mathcal{N}^{c}_{\emph{v}_{\emph{i}}}}\alpha_{\emph{v}_{\emph{i}}c_{\emph{x}}}^{(m-1)} \textbf{h}_{\emph{c}_\emph{x}}^{(\emph{m}-1)}
\end{equation}

Finally, we synthesize both types of contextual dependencies to update the final embedding of the item $v_i$,
\begin{equation}
	\textbf{h}_{v_i} = \Tilde{\textbf{h}}_{v_i} + \delta \textbf{h}_{\emph{v}_{\emph{i}}^{c}} 
\end{equation}
where $\delta$ serves as a trade-off factor. Based on this, CGGNN captures contextual dependencies from neighboring entities and categories simultaneously as well as generates a high-order representation for items.

\subsection{Dual-agent Reinforcement Learning}


Typically, existing RL-based explainable recommendation methods over KGs employ single-agent RL frameworks to conduct explainable recommendations over short paths \cite{PGPR} \cite{ADAC}. However, the insufficient feedback of reward signals and the large action space constrain the performance of these methods as path length increases. To address the above problems, we propose a novel dual-agent RL framework to efficiently find items over long recommendation paths. The main innovations of this component are twofold. (1) The novel dual-agent RL framework overcomes the inherent technical limitations of existing RL-based methods that rely on short paths. (2) Our RL framework with a collaborative paradigm provides sufficient reward signals and reduces action spaces, offering a new perspective for RL-based explainable recommendations.


Technically, DARL contains two agents: the category agent and the entity agent. The former quickly traverses over $\mathcal{G}^c$ and is committed to finding the item-category where the target item is located, offering milestone-like category-level guidance to the latter to narrow the action spaces. The latter efficiently walks over various types of entities within the categories based on milestone-like hints to identify suitable items for the user. Moreover, the dual agents share historical paths and rewards with each other through shared policy networks and a collaborative reward mechanism. Consequently, DARL comprehensively leverages both global and local perspectives (i.e., category and entity views)  to efficiently optimize the recommendation decision-making process.


\subsubsection{Category Agent}

The category agent considers an item-category as an abstract node and selects different nodes as subsequent actions until reaching the target category. Note that, the target category is where the target item lies, while the initial category contains the item directly associated with the user. Category agent interacts with categories in $\mathcal{G}^c$ through a 4-tuple of MDP (State, Action, Transition, Reward). 


Specifically, each state at recommendation step $l$ is represented as $s_l^c$ = ($u$, $c_s$, $c_l$), where $c_l$ $\in$ $\mathcal{G}^c$ denotes the current category, and $c_s$ indicates the initial category. The state space $\mathcal{S}^c$ is conceptualized as a collection of states. In addition, an action space $\mathcal{A}^c$ contains the set of valid actions $A_l^c$ at recommendation step $l$, which drives the category agent to move from the current category to the next. Formally, $A_l^c$ is formulated as $A_l^c$ = \{$c'$, $\mid$ ($c_l$, $c'$) $\in$ $\mathcal{G}^c$\}. Considering that the category-level recommendation path is shorter than that of the entity-level, the self-loop action is set to prevent infinite unrolling of paths and synchronize with the entity agent. 
The next state $s_{l+1}^c$ is updated by the previous state $s_{l}^c$ according to the transition function  $\mathcal{P}^c$: $\mathcal{S}^c$ $\times$ $\mathcal{A}^c$ $\rightarrow$ $\mathcal{S}^c$. As a crucial element in the RL framework, the terminal reward function $\Tilde{R}^c$ of the category agent is defined as $\Tilde{R}^c$ = $\mathbbm{1}$($c_L$), where $\mathbbm{1}$ is an indicator function, i.e., this agent receives a reward signal of 1 if it reaches the target category. Otherwise, the value of $\Tilde{R}^c$ is 0. Notably, the final reward $R^{c}$ of the category agent is defined in Eq. (20).

\subsubsection{Entity Agent}
Similar to the single-agent explainable recommendation methods \cite{PGPR} \cite{ADAC} \cite{UCPR},  the goal of the entity agent is to reach the appropriate items by traversing $\mathcal{G}$. Formally, each state at the recommendation step $l$ is denoted as $s_l^e$ = ($u$, $e_l$), where $e_l$ $\in$ $\mathcal{E}$, and $\mathcal{U}$, $\mathcal{V}$, $\mathcal{F}$, $\mathcal{B}$ $\subseteq$ $\mathcal{E}$. Note that, if the current entity type is an item, its representation $\textbf{h}_{v}$ is derived from CGGNN, otherwise, the entity representation is generated by TransE. Furthermore, the terminal state is $s_L^e$ = ($u$, $v_L$). For state $s_l^e$ at the step  $l$, the valid action of our entity agent is formulated as $A^e_l$ = \{($r'$, $e'$) $\mid$ ($e_l$, $r'$, $e'$) $\in$ $\mathcal{G}$\}, where $A^e_l$ $\in$ $\mathcal{A}^e$.  In addition, the transition function $\mathcal{P}^e$ of the entity agent defines the process from a current state to the next state. Within a maximum number of recommendation steps $L$, the reward is defined as $\Tilde{R}^e$ = $\mathbbm{1}_{\emph{V}_u}$($e_L$). The value of $\mathbbm{1}_{\emph{V}_u}$($e_L$) is 1 when $e_L$ $\in$ $\mathcal{V}_u$ and 0 when $e_L$ $\not \in$ $\mathcal{V}_u$. Note that, the final reward $R^{e}$ of the entity agent is defined in Eq. (21).


\subsubsection{Shared Policy Networks}
The proposed shared policy networks aim to learn path-finding policies to maximize the expected cumulative reward during recommendation. Specifically, we design two networks $\pi_{\theta}^c$ and $\pi_{\theta}^e$ to model the behavioral policies of category and entity agents, enabling the networks to achieve optimization goals collaboratively. 
Following \cite{EX7}, we first use LSTM to encode the searched trajectories and then calculate the historical sequence of the entity and category  agents. For the category agent, the historical information at step $l$ is $y^c_l$ = \{$u_0$, $c_0$, $c_1$,...,$c_l$\} $\in$ $ \mathcal{Y}^c$. Similarly, the entity agent's historical information is defined as $y^e_l$ = \{$u_0$, $r_0$, $e_0$,..., $r_l$, $e_l$\} $\in$ $ \mathcal{Y}^e$. This historical information can be encoded by two separate LSTMs, 
\begin{equation}
	\textbf{y}_{0}^{c} =  \mathrm{LSTM}_{c}(\textbf{0},[\textbf{u}_0; \textbf{c}_0]), \textbf{y}_{0}^{e} = \mathrm{LSTM}_{\emph{e}}(\textbf{0},[\textbf{u}_0; \textbf{h}_{r_0}; \textbf{h}_{e_0}])
\end{equation}
\begin{equation}
	\textbf{y}_{l}^{c} = \mathrm{LSTM}_{\emph{c}}(\textbf{W}^c[\textbf{y}_{\emph{l}-1}^c; \textbf{y}_{\emph{l}-1}^e],\textbf{c}_{\emph{l}-1}), \emph{l} \textgreater 0,
\end{equation}
\begin{equation}
	\textbf{y}_{l}^{e} = \mathrm{LSTM}_{\emph{e}}(\textbf{W}^e[\textbf{y}_{l-1}^e; \textbf{y}_{l-1}^c],[\textbf{h}_{r_{l-1}}; \textbf{h}_{e_{l-1}}]), \emph{l} \textgreater 0,
\end{equation}
where $\textbf{W}^e$ and $\textbf{W}^c$ are transformation matrices that prevent the exponential increase of dimensions. 
As mentioned above, if the current entity $e_l$ is an item, its representation $\textbf{h}_{v_l}$ is from CGGNN. For simplicity of representation, the representations of the different types of entities are uniformly denoted as $\textbf{h}_e$.
After modifying the internal structure of the LSTM to complete the history shaping, each hidden state of the agent ensures that necessary path information is shared with another agent to achieve coordinated action selection and the reduction of action spaces.
The set of possible actions $A^c$ and $A^e$ are encoded by stacking the embeddings of all actions: $\textbf{A}^c$ $\in$ $\mathbb{R}^{\left| A^c \right| \times 7d}$, $\textbf{A}^e$ $\in$ $\mathbb{R}^{\left| A^e \right| \times 7d}$. The shared policy networks are defined as follows,
\begin{equation}
    \emph{$\pi$}_\theta^{c}(\emph{a}_l^c|\emph{s}_l^c) =  \mathrm {softmax} (\textbf{A}_{l}^c\times \textbf{W}_{2}^c\text{ReLu}(\textbf{W}_{1}^c[\textbf{u}_0; \textbf{c}_l;\textbf{y}_l^c]))
\end{equation}
\begin{equation}
    \emph{$\pi$}_\theta^{e}(\emph{a}_l^e|\emph{s}_l^e) = \mathrm { softmax} (\textbf{A}_{l}^e\times \textbf{W}_{2}^e\text{ReLu}(\textbf{W}_{1}^e[\textbf{h}_{e_l};\textbf{h}_{r_l};\textbf{y}_l^e]))
\end{equation}

Based on this, shared policy networks can  simultaneously calculate  category-level and entity-level information to predict the next action for dual agents.

\subsubsection{Collaborative Reward Mechanism}

The default terminal rewards of the dual agents only consider whether they can achieve the pre-defined goal, neglecting the guidance and consistency of paths. This exacerbates the negative impact of sparse reward dilemma  and large action spaces on the recommendation performance as the number of hops increases. Specifically, (1) \emph{Guidance}: the entity agent requires stage-wise guidance from the category agent to reduce the action space and enhance efficiency. (2) \emph{Consistency}: the category agent traverses over  $\mathcal{G}_c$ with a densely connected structure, resulting in more diverse paths and difficulty maintaining consistency with entity level paths. However, the overlap of the paths of dual agents can enhance the effectiveness of their decision-making \cite{path1zx}. Based on this, we propose a collaborative reward mechanism in this subsection. It models the causal influence between dual agents by generating partner rewards, and then combines their respective terminal rewards to form final reward functions. The innovation of this collaborative reward mechanism lies in two key aspects: (1) allowing the category agent to provide efficient path-finding guidance for the entity agent by evaluating different actions \cite{EX6}; (2) constraining  the category agent to generate category-level movement trajectories consistent with entity-level paths.

Counterfactual actions are introduced to model the path guidance from  various  categories. Specifically, we first define a joint probability for the potential actions of the entity agent, denoted as  $p(a_l^e|a_l^c,s_l^e)$. Then, an optional action $\Tilde{a}_l^c$ within the action set $A_l^c$ is introduced to replace the intervention of action $a_l^c$ during the recommendation process. Next, we calculate the conditional probability distribution $p(a_l^e|\Tilde{a}_l^c,s_l^e)$ based on the action and state spaces. Essentially, the category agent spontaneously poses a retrospective question: how does the introduction of other categories affect the path-finding of the entity agent? 
Following this, the marginal probability $p(a_l^e|s_l^e)$ = $\sum_{\Tilde{a}_l^c} p(a_l^e|\Tilde{a}_l^c,s_l^e)p(\Tilde{a}_l^c|s_l^e)$ without considering $a_l^c$ is obtained by averaging the resulting probability distribution of the entity agent. Noted here, 
the discrepancy between $p(a_l^e|s_l^e)$ and $p(a_l^e|a_l^c,s_l^e)$ 
 quantifies the causal impact of the category on the entity agent. 
Therefore, we utilize the Kullback-Leibler Divergence to formalize  the above process and define partner reward $R^{p_c}$ from the category agent to the entity agent at the recommendation step $l$.
\begin{equation}
\begin{split}
\large
   \varphi_{l}=D_{KL}[p(a_l^e|a_l^c,s_l^e)\| \sum_{\Tilde{a}_l^c} p(a_l^e|\Tilde{a}_l^c,s_l^e)p(\Tilde{a}_l^c|s_l^e)] \\
   =D_{KL}[p(a_l^e|a_l^c,s_l^e)\|p(a_l^e|s_l^e)]
   \end{split}
\end{equation}
\begin{equation}
R^{p_c}_l = \frac{1}{1+\rm{exp}(-\varphi_{\emph{l}})}
\end{equation}

In addition, the currently accessed category must be semantically similar to the item corresponding to the step $l$. We measure the state similarity of two agents through the cosine similarity function. 
\begin{equation}
R^{p_e}_l= \frac{{\textbf{\textit{s}}_l^c \cdot \textbf{\textit{s}}_l^e}}{{\|\textbf{\textit{s}}_l^c\| \cdot \|\textbf{\textit{s}}_l^e\|}}
\end{equation}
where $R^{p_e}_l$ is a partner reward to control the intensity of path consistency.

The final rewards of the category and entity agents at the step $l$ are defined as $R^{c}(s_l^c)$ and $R^{e}(s_l^e)$, respectively. 
\begin{equation}
   R^{c}(s_l^c)=\Tilde{R}^c(s_L^c) + \alpha_{p_e}R^{p_e}_l, l \in [1,\emph{L}]
\end{equation}
\begin{equation}
   R^{e}(s_l^e)=\Tilde{R}^e(s_L^e)+ \alpha_{p_c} R^{p_c}_l, l \in [1,\emph{L}]
\end{equation}
where $\alpha_{p_e}$ and $\alpha_{p_c}$ denote reward discount factors. Note that, dual agents can obtain the sufficient reward at every step by integrating terminal and partner rewards, which ensures performance stability \cite{DREAM}\cite{MARLsurvey1}.

Finally, we update the model parameters by using the REINFORCE algorithm \cite{REINFORCE} with the stochastic gradient. The dual-agent RL framework fully utilizes shared policy networks and the collaborative reward mechanism to narrow the action space and eliminate the sparse reward dilemma, thereby enhancing the model's effectiveness and efficiency over KGs. 

\section{Experiments}
In this section, we first introduce the experimental setup and baselines, followed by conducting experiments to investigate the following research questions (RQs). \textbf{RQ1}: Does the recommendation accuracy of CADRL consistently outperform state-of-the-art (SOTA) baselines on real-world datasets? \textbf{RQ2}: Is CADRL more efficient than existing RL-based recommendation methods over KGs? \textbf{RQ3}: How do the main components of CADRL affect recommendation performance?  \textbf{RQ4}: What are the contributions  of the  technological modules in CADRL to recommendation performance? \textbf{RQ5}: What role does path length play in RL-based recommendation models?  \textbf{RQ6}: What impact do different key hyper-parameters have on the performance of CADRL? \textbf{RQ7}: How do we evaluate the explainability of our proposed CADRL?

\begin{table*}
  \centering
  \caption{Comparison of Recommendation Accuracy  on Amazon datasets.}
  \vspace{-0.2cm}
  \setlength{\tabcolsep}{2.5mm}{
   \begin{tabular}{l|cccc|cccc|cccc}
   	\toprule
    &  \multicolumn{4}{c|}{Clothing} & \multicolumn{4}{c|}{Cell Phones}& \multicolumn{4}{c}{Beauty} \\ 
    
    Model         & NDCG   & Recall    &HR  &Prec.  & NDCG   & Recall    &HR  &Prec.  & NDCG   & Recall    &HR  &Prec.  \\ 
    \hline
    CKE     &1.207 &2.026  &3.424  &0.312 &3.421 &5.997 &9.269 &0.917 &3.087 &4.831  &9.275  &1.149\\
    KGAT & 2.439 & 4.142  & 6.027   & 0.603   &4.345 &7.686  &10.946  &1.092 &5.023 &8.378 &14.181 &1.531\\
    DeepCoNN  &1.066  &1.885  &2.665   &0.186   &3.164 &5.478  &8.406  &0.854 &2.732 &4.561 &8.231 &1.013\\
    RippleNet   &1.749  &3.188  &4.826   &0.493   &4.135 &6.656  &9.681  &0.937 &4.368 &6.894 &12.180 &1.417\\

   RuleRec    &1.817  &3.205  &4.915  &0.504   &4.332 &6.779  &9.782  &1.022 &4.472 &6.932 &12.274 &1.479\\
       HeteroEmbed & 2.514 & 4.482  & 6.461   & 0.621   &4.585 &7.913  &11.481  &1.124 &5.221 &8.741 &14.413  &1.612     \\
PGPR  &2.362 &3.963  &5.852  &0.596   &4.340 &7.248 &10.246  &1.089 &4.618 &7.538 &12.814 &1.492 \\
    ReMR  &2.534  &4.496  &6.517   &0.635     &4.601   &7.963   &11.597    &1.188 &5.279 &8.773 &14.686 &1.647\\

    ADAC   &2.612  &4.511  &6.635   &0.664  &4.627  &7.988  &11.641 &1.201 &5.321 &8.822 &14.902 &1.687 \\
    
    INFER  &2.726  &4.583  &6.735  &0.701    &4.704   &8.049  &11.717   &1.273 &5.388 &8.873 &14.985 &1.736\\

    CogER   &2.775  &4.613  &6.804  &0.773    &4.932   &8.336   &12.243    &1.306 &5.473 &9.053 &15.214 &1.826\\

    CAFE  &2.802  &4.657  &6.875  &0.798 &5.313  &8.886   &13.031   &1.372 &6.163 &9.716 &15.827 &2.352\\

        UCPR  &{\underline {2.978}}  &{\underline {4.969}}  &{\underline {7.343}}  &{\underline {0.863}}    & {\underline {5.542}} &{\underline {9.454}} &{\underline {13.721}} &{\underline {1.463}} &{\underline {6.533}} &{\underline {10.431}} &{\underline {16.919}} &{\underline {2.592}}\\



    {\bf CADRL }  &{\bf3.259}  &{\bf5.368}   &{\bf7.963}  & {\bf0.937}  & {\bf 6.397} & {\bf10.752}  & {\bf15.667}  & {\bf1.691} & {\bf7.738}  & {\bf12.462}  & {\bf20.429} &{\bf3.134}\\
    \hline
        Improv.    &9.44\%   & 8.03\%  &8.44\% &  8.57\% &15.43\%  &13.73\% &14.18\% & 15.58\%  &18.44\% &19.47\% &20.75\% &20.91\% \\
   	\bottomrule
   \end{tabular}
  }
  \label{Table2}
 \end{table*}

\begin{table}
	\centering
	\caption{Statistics of the Experimental Datasets}
 \vspace{-0.2cm}
	\begin{tabular}{lllll}
		\hline
		Dataset     & \#Beauty  & \#Cell Phones        & \#Clothing  &   \\
		\hline
		\#Users       & 22,363  &27,879    & 39,387 \\
		\#Items      & 12,101  & 10,429      & 23,033 \\
  \#Entities      &59,105  &61,756      &84,968\\
		\#Interactions    & 127,635  &141,076    & 181,295 \\
            \#Triplets    & 1,903,246  &1,253,283     & 2,745,308\\
		\hline
	\end{tabular}
	\label{Table1}
\end{table}
 
\subsection{Experimental Setup}
\subsubsection{Datasets} We report the recommendation results of CADRL on  public real-world datasets$\footnote{http://jmcauley.ucsd.edu/data/amazon}$, i.e., Beauty, Clothing and Cell Phones datasets from the Amazon e-commerce data collection \cite{data1}\cite{data2}. Following the data processing in studies \cite{ADAC} \cite{CogER}, each dataset is constructed as an individual KG containing 4 types of entities (i.e., \emph{User}, \emph{Item}, \emph{Brand} and \emph{Feature}) and 14 types of relations (\emph{Purchase}, \emph{Mention}, \emph{Described\_by}, \emph{Produced\_by}, \emph{Also\_bought}, \emph{Also\_viewed}, \emph{Bought\_together}, and their 7 corresponding inverse relations). Note that, categories belong to the top-level ontology in KGs and are not at the same level as entities \cite{ontology2} \cite{ontology1}. Thus, unlike prior work \cite{PGPR} that treats a category as an entity, our work regards the concept of a category as an abstract collection of items. Following the principle of data partitioning in previous studies \cite{PGPR} \cite{UCPR} \cite{ReMR},  we randomly select 70\% of user purchases (interactions) as the training data and consider the remaining 30\% as the test data. The statistics of all datasets are summarized in Table \ref{Table1}.

\subsubsection{Evaluation Protocol}Following previous studies \cite{ADAC} \cite{ReMR}, we evaluate the performance of  recommendation methods using widely-used metrics,  including Precision (Prec.), Recall, Normalized Discounted Cumulative Gain (NDCG), and Hit Ratio (HR). The above indicator scores positively correlated with  recommendation performance. Notably,  we calculate the ranking metrics based on the top-10 recommended items for each user in the test set.
 
\subsubsection{Hyper-Parameters}
Similar to existing methods \cite{ADAC} \cite{CogER}, CADRL is trained for 50 epochs using Adam optimizer  with a learning rate of 0.0001 across  all datasets. In addition, the embedding size for all entities, relations, and categories is set to 100. The maximum sizes for  $\mathcal{A}^c$ and $\mathcal{A}^e$ are set to 10 and 50, respectively. 
Amazon’s metadata has defined item’s category details \cite{PGPR}, which determines the number of categories obtained by grouping the items to be 248, 206, and 1,193 on Beauty, Cell Phones, and Clothing, respectively. Moreover, the number of GNN layers $k$, $m$, reward discount factors $\alpha_{p_e}$, $\alpha_{p_c}$ and the maximum recommendation length $L$ are set to [3, 2,  0.6, 0.5, 6] and [3, 2,  0.4, 0.5, 6] on Beauty and Cell Phones, respectively, while they are set to [3, 2,  0.4, 0.4, 7] on the Clothing dataset, respectively. The trade-off factors $\delta$ are set to 0.3, 0.4, and 0.4 on Clothing, Cell Phones, and Beauty, respectively. These parameters represent the optimal values obtained through parameter tuning.

\subsection{Baselines}
To investigate the performance of CADRL, we use the following four categories of recommendation models as baselines in our experimental comparisons. 
\begin{itemize}
\item KG embedding models: CKE \cite{CKE} extracts structural representations of items by integrating  the heterogeneous information from KGs, while KGAT \cite{KGAT} refines embeddings through the GAT mechanism to capture high-order connectivities within KGs.

\item Neural networks-based models: DeepCoNN  \cite{DeepCoNN} utilizes  cooperative deep neural networks  to encode both users and items for rating prediction. In addition, RippleNet \cite{Ripplenet} iteratively expands the user's potential interests along relations in an end-to-end manner, which can be used to calculate the final purchase probability for users. 

\item Traditional path models: RuleRec \cite{KGRSrules} constructs a rule-guided mechanism to predict items based on the meta-knowledge of recommendation paths. HeteroEmbed \cite{HeteroEmbed} is a path-based SOTA recommendation model that traverses potential recommendation paths over KGs by learning user-item connective patterns. 

\item RL-based models:  This group of models is closely related to our work and is elaborated upon in Section II. We compare not only the well-known PGPR \cite{PGPR}, CAFE \cite{CAFE}, and ADAC \cite{ADAC}, but also the latest models in this domain, such as INFER \cite{INFER}, ReMR \cite{ReMR}, UCPR \cite{UCPR}, and CogER \cite{CogER}. Notably, UCPR \cite{UCPR} is a SOTA RL-based method that demonstrates strong performance over KGs.

\end{itemize}

\begin{table*}

\small
  \centering
  \begin{threeparttable}
  \caption{The Computational Cost with Running Time 
   Regarding Recommendation and Path Finding.}

\label{Table3}
    \begin{tabular}{ccccccccc}
    \toprule
    \multirow{2}{*}{Time(s)}&
    \multicolumn{2}{c}{Clothing}&\multicolumn{2}{c}{Cell Phones}&\multicolumn{2}{c}{Beauty}\cr
    \cmidrule(lr){2-3} \cmidrule(lr){4-5} \cmidrule(lr){6-7}
    &Rec. (1k users)      & Find (10k paths)  &Rec. (1k users)      & Find (10k paths)
    &Rec. (1k users)      & Find (10k paths)\cr
    \hline
    PGPR & 217.232 $\pm$ 4.127      & 18.635 $\pm$ 0.402 & 262.218 $\pm$ 4.327 & 21.720 $\pm$ 0.472 &274.501 $\pm$ 5.741 &22.108 $\pm$ 0.512\cr
    HeteroEmbed &50.266 $\pm$ 1.318    &  15.827 $\pm$  0.307  &44.246 $\pm$ 1.024  & 16.682 $\pm$ 0.401  &46.734 $\pm$ 1.166 &18.682 $\pm$ 0.396\cr
    UCPR &  30.635 $\pm$ 1.134     & 14.293 $\pm$ 0.295   & 29.578 $\pm$ 1.025       &  14.981 $\pm$ 0.302  &33.775 $\pm$ 1.146 &14.473 $\pm$ 0.302 \cr
    CAFE &  22.043 $\pm$ 1.052    & 13.114 $ \pm$ 0.268  &  21.880 $\pm$ 1.022      & 13.381 $\pm$  0.337 &24.761 $\pm$ 1.061 &12.533 $\pm$ 0.342\cr
    CADRL &{\bf20.625 $\pm$ 1.025}    & {\bf12.205 $\pm$ 0.113}   & {\bf18.386 $\pm$ 1.017}       &  {\bf11.334 $\pm$ 0.251}  &{\bf21.592 $\pm$ 1.028} &{\bf10.516 $\pm$ 0.235} \cr
    
  	\bottomrule
    \end{tabular}
    \end{threeparttable}
\end{table*}

\begin{table*}
	\centering
	\caption{Ablation on Different Components on Real-world Datasets.}
	\setlength{\tabcolsep}{2.5mm}{
		\begin{tabular}{c|cccc|cccc|cccc}
			\toprule
			&  \multicolumn{4}{c|}{Clothing} & \multicolumn{4}{c|}{Cell Phones} & \multicolumn{4}{c}{Beauty} \\ 
			Model         & NDCG &Recall &HR &Prec.     & NDCG &Recall &HR &Prec.   & NDCG &Recall &HR &Prec.  \\ \midrule

   	CADRL w/o  DARL  & 2.847    & 4.720 & 6.912 &0.843 & 5.397  & 8.915 & 13.483 &1.397  & 6.209    &9.810 & 16.102  &2.389  \\
      
    CADRL w/o   CGGNN & {\underline {3.122}}      & {\underline {5.242}} & {\underline {7.746}} & {\underline {0.913}} & {\underline {6.129}} & {\underline {10.243}}  &{\underline {14.956}} &{\underline {1.539}} &{\underline {7.316}}    & {\underline {11.583}} &{\underline {19.254 }}   &{\underline {2.955}}\\

     CADRL  &{\bf3.259}  &{\bf5.368}   &{\bf7.963}  & {\bf0.937}  & {\bf 6.397} & {\bf10.752}  & {\bf15.667}  & {\bf1.691} & {\bf7.738}  & {\bf12.462}  & {\bf20.429} &{\bf3.134}\\
			\bottomrule

		\end{tabular}
	}
	\label{table4}
\end{table*}

\subsection{Recommendation Accuracy (RQ1)}

In this subsection, the results of recommendation accuracy are illustrated in Table~\ref{Table2} (all scores are reported as percentages). The most competitive baseline results are underlined, while the best results are highlighted in bold.  Specifically, we provide the following insightful analyses. (1) 
CADRL consistently surpasses existing baselines across all metrics on real-world datasets. This performance advantage primarily arises from two key aspects. Firstly, the high-order item representation of CADRL fully captures contextual dependencies and models latent semantic information. Secondly, the dual-agent RL  framework of CADRL collaboratively mines long-path connection patterns and  potential path combinations to adaptively obtain the optimal recommendation policy. (2) RL-based methods generally outperform other recommendation baselines. For example, compared to the path-based SOTA model HeteroEmbed, UCPR exhibits  NDCG performance gains of 18.46\%, 20.87\%, and 25.13\% on Clothing, Cell Phones, and Beauty, respectively. This expected result demonstrates that the dynamic decision-making capability of these RL-based methods  significantly enhances recommendation accuracy. (3) Compared to other SOTA recommendation methods, CADRL demonstrates a smaller performance improvement on the Clothing dataset relative to other datasets.  One potential reason is that the categories 
 within the Clothing domain contain a smaller number of items (i.e., approximately 19.31, 48.79, and 50.63 items per category in Clothing, Beauty, and Cell Phones, respectively). This sparsity within categories weakens the effectiveness of CADRL in leveraging categories as optimal metadata for inferring user preferences.



\subsection{Efficiency Study (RQ2)}
Following previous work \cite{CAFE}, we investigate  efficiency by comparing the time consumption of various methods in recommendation and pathfinding. To ensure a fair comparison, we use only path-based and RL-based methods with their optimal parameter settings, including the SOTA path-based method HeteroEmbed and the 3-hop RL-based explainable recommendation methods PGPR, UCPR, and CAFE. Specifically, the efficiency of these methods is evaluated using two metrics: (1) the empirical running time to infer suitable items for 1k users; and (2) the time consumed to generate 10k recommendation paths.  The results are reported in Table \ref{Table3}.

We first observe that CADRL exhibits the fastest performance across all test cases. This is because CADRL reduces the action space through the collaboration of dual agents at both the entity and category levels.  Furthermore, the additional path sorting process increases the running time of PGPR, resulting in its inefficiency in both recommendation and pathfinding. Different from Table~\ref{Table2}, UCPR does not show the strongest efficiency advantage among the baselines. A potential explanation is that  the modeling of demand portfolios at each step increases UCPR's time consumption. Compared with the most efficient baseline CAFE, CADRL has an overall time advantage on all datasets.  Also, the maximum and average action space of CAFE are $\mathcal{O}$($\left| \mathcal{E} \right|$$\times$$\left|L\right|$) and $\mathcal{O}$($\left| \mathcal{A}^e \right|$$\times$$\left|L\right|$) while those of CADRL are reduced to $\mathcal{O}$($\frac{\left| \mathcal{E} \right|}{\left| C \right|}$$\times$$\left|L\right|$) and $\mathcal{O}$($\frac{\left|\mathcal{A}^e\right|}{\left|\mathcal{A}^c\right|}$$\times$$\left|L\right|$), where $L$ $\ll$ $\left| C \right|$ and $\left|\mathcal{A}^c\right|$. This explains why CADRL is more efficient than SOTA baselines even though it uses more hops. In brief, Table \ref{Table3}, together with Table~\ref{Table2}, reveals that CADRL’s recommendation efficiency and effectiveness surpass those of existing SOTA methods across varying-scale datasets.





\subsection{Ablation Study (RQ3)}

The primary objective of the ablation study is to assess the contributions of two components (i.e., CGGNN and DARL) in CADRL. Given the necessity of the entity agent for explainable recommendations, the first variant \textbf{CADRL w/o DARL} only retains the structure of a single agent and masks the participation of the category agent in the policy networks and reward functions. Compared to existing RL-based explainable recommendation methods, the RL design that only includes binary terminal rewards of CADRL w/o DARL is the simplest and most basic. Additionally, \textbf{CADRL w/o CGGNN} removes CGGNN and adopts solely the static representation generated by TransE. Table \ref{table4} presents the experimental results of the CADRL and its variants.

\begin{figure*}
	\centering
	\subfigure[NDCG]
	{
		\centering
		\includegraphics[width=0.231\linewidth,height=2.7cm]{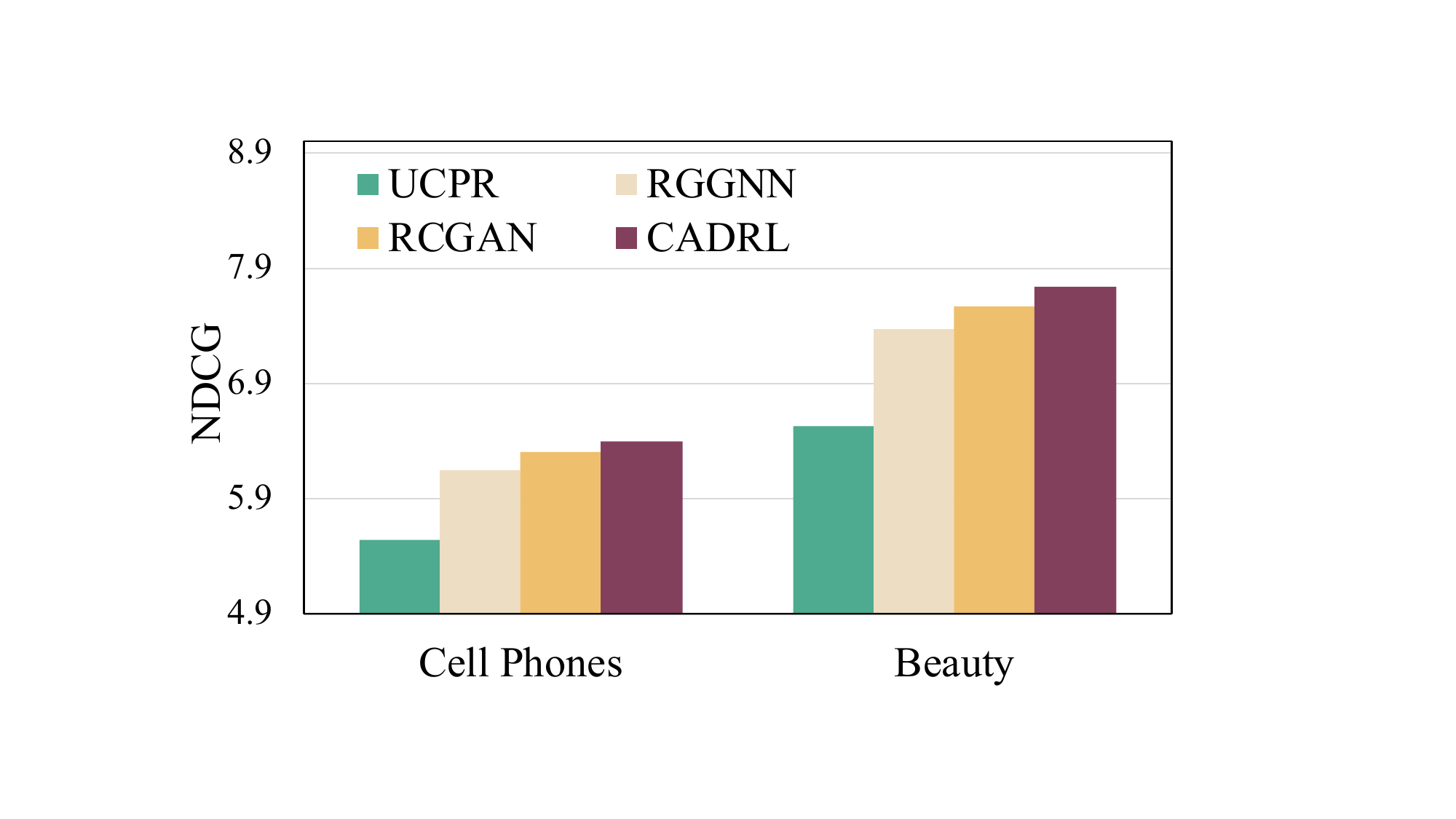}
	}
	\subfigure[Recall]
	{
		\centering
		\includegraphics[width=0.231\linewidth,height=2.7cm]{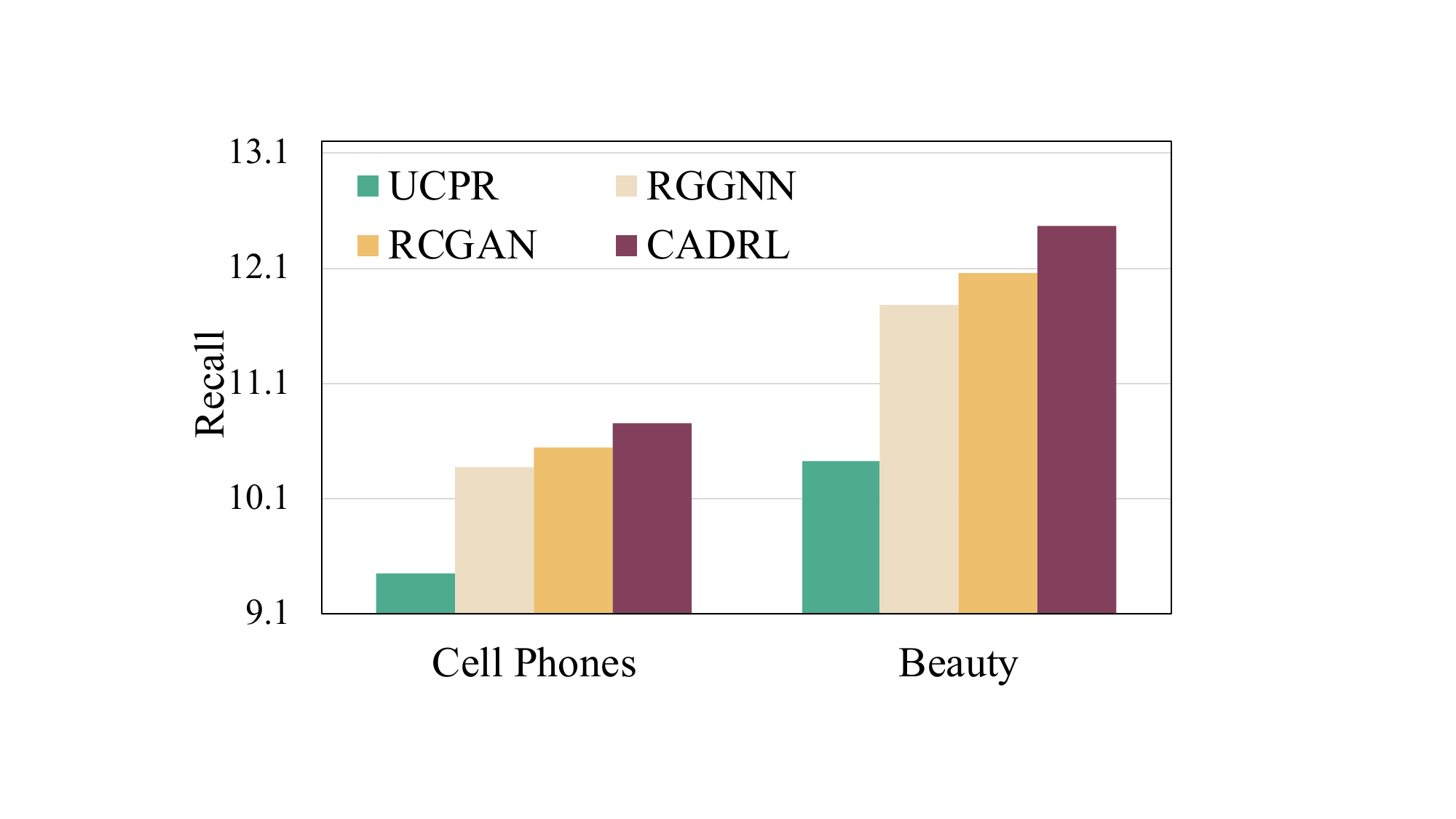}
	}
 \subfigure[HR]
	{
		\centering
		\includegraphics[width=0.231\linewidth,height=2.7cm]{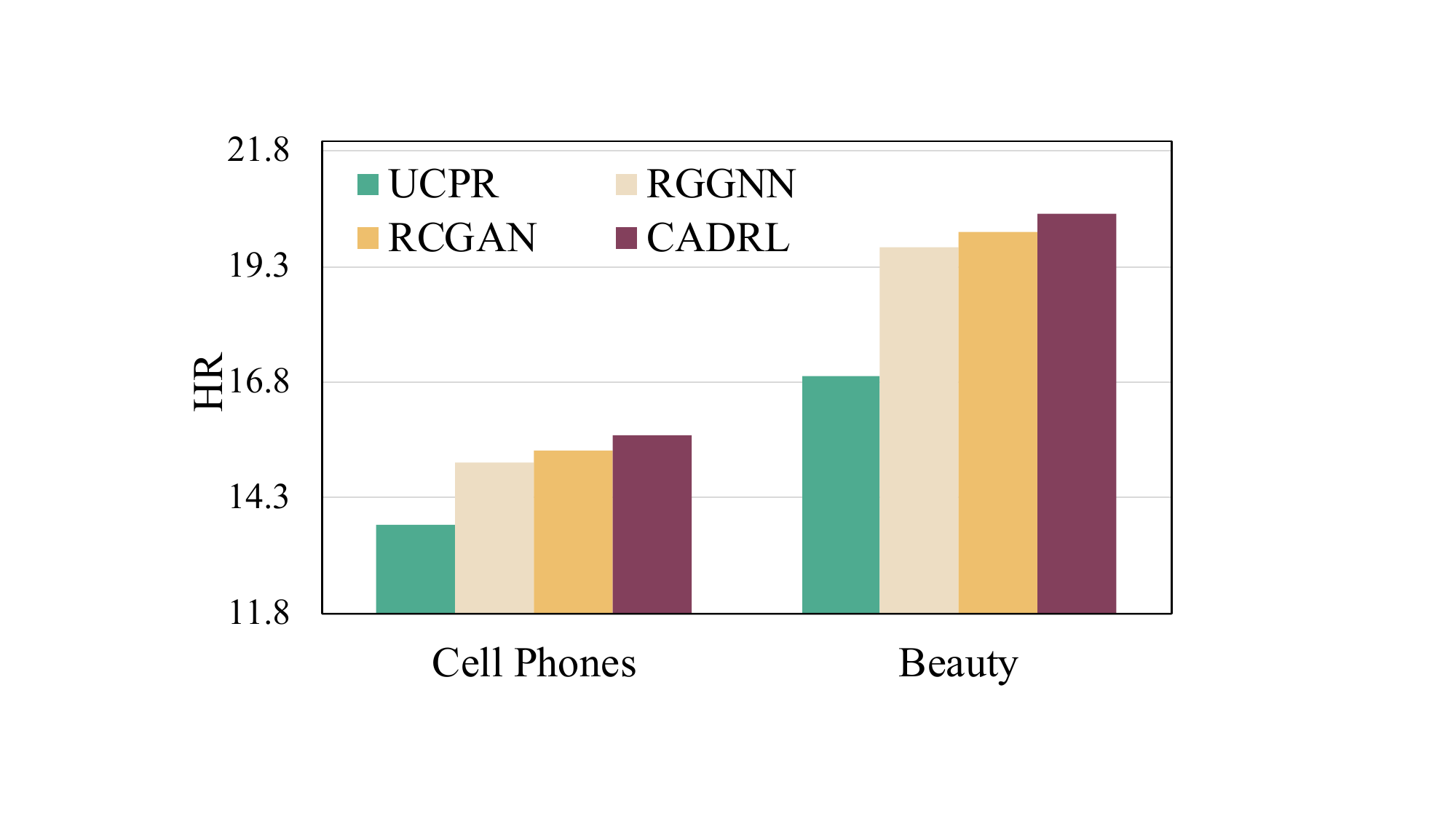}
	}
  \subfigure[Precision]
	{
		\centering
		\includegraphics[width=0.231\linewidth,height=2.7cm]{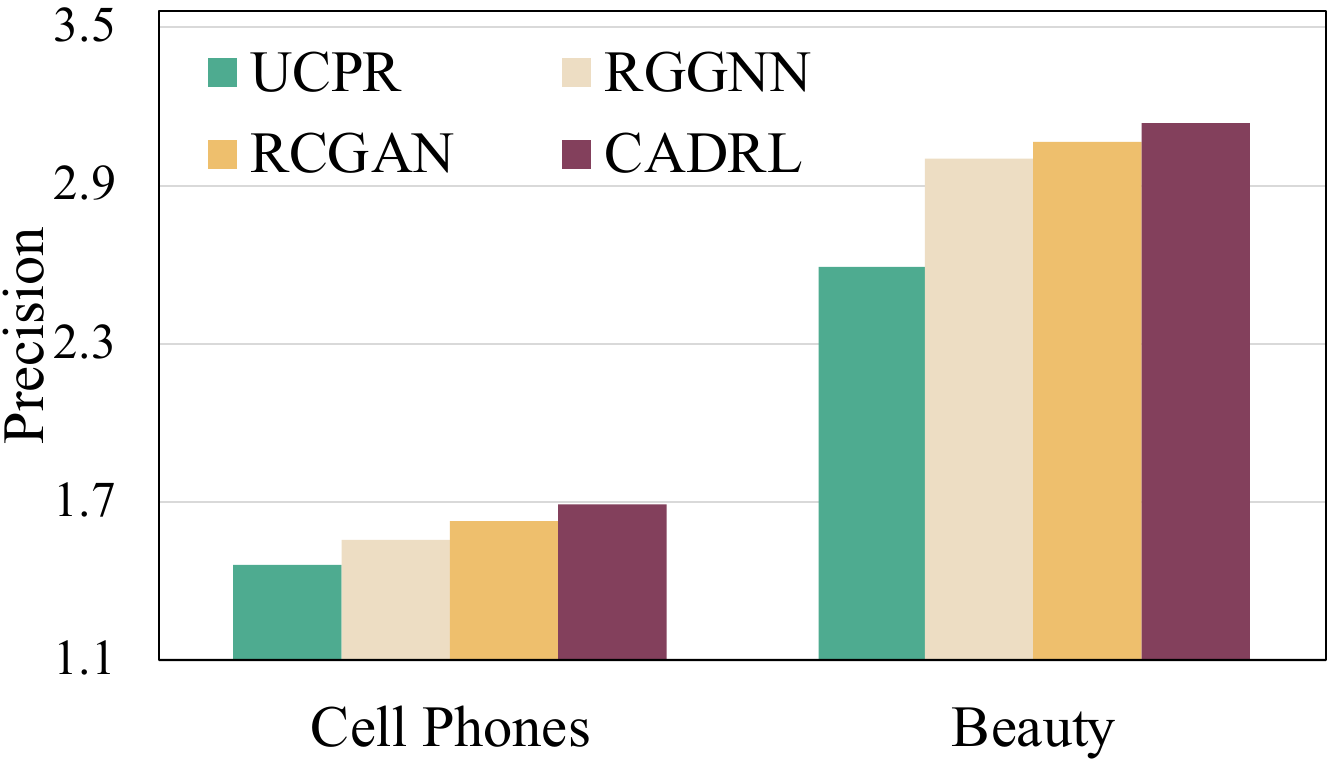}
	}
	\caption{Effectiveness  analysis of different modules in CGGNN on different datasets.
 }
	\label{fig3}
\end{figure*}

We have the following observations. (1) Undoubtedly,  CADRL w/o DARL without the involvement of the dual-agent collaboration mechanism has the lowest performance among the above methods. One potential explanation is that it lacks stage-wise guidance from the category agent and easily suffers from sparse reward dilemmas \cite{MARLsurvey}. 
(2) Although the performance of CADRL w/o DARL is diminished due to the absence of the DARL component, it still surpasses some existing baselines. For instance, the NDCG of CADRL w/o DARL on Clothing, Cell Phones and Beauty are respectively 2.6\%, 9.4\% and 13.4\% higher than those of CogER.  This is because CGGNN in CADRL w/o DARL can generate high-order item representations encompassing context dependencies, thereby enhancing the accuracy of recommendations. (3) The recommendation performance of CADRL w/o CGGNN is generally higher than that of CADRL w/o DARL but lower than that of CADRL on all datasets. This experimental result further demonstrates the effectiveness of CGGNN and implies that  the DARL component contributes more significantly to the performance of CADRL than the CGGNN component. Notably,  with the exclusion of the dual-agent collaboration mechanism, the reward function of CADRL w/o DARL consists solely of basic terminal rewards, which do not address the sparse reward dilemma.  To report in detail the impact of the reward function as well as the other modules on the recommendation effectiveness of CADRL, we perform the effectiveness analysis in the following subsection.



\subsection{Effectiveness Analysis (RQ4)}
To investigate the impact of the modules 
 on the aforementioned components, we introduce several variants of the method by removing specific technology modules to conduct an effectiveness analysis. Specifically, we first study the recommendation performance of two modules within CGGNN, namely the  \emph{Gated Graph Neural Network} (GGNN) and the \emph{Category-aware Graph Attention Network} (CGAN).  Subsequently, we explore the effectiveness of the \emph{Shared Policy Networks} (SPN) and the \emph{Collaborative Reward Mechanism} (CRM) within DARL. Note that, the experimental results in this subsection are conducted on Beauty and Cell Phones to observe obvious numerical differences. 

\subsubsection{The contribution of various  modules in CGGNN}
GGNN and CGAN capture contextual dependencies at the entity level and category level, respectively, aiming to generate high-order representations of items. To verify the effectiveness of these modules, we design the following variant models: (1) RGGNN: this variant removes only the GGNN from CADRL, preventing the capture of entity-level contextual dependencies and relying solely on static representations by TransE for items. (2) RCGAN: only the CGAN module is removed from CADRL, leaving only GGNN in this variant model to produce item representations. 

The results are presented in Fig. \ref{fig3}.  There are three main discoveries. (1) The variant models outperform the SOTA baseline UCPR in all metrics, which illustrates the recommendation effectiveness of both modules in CGGNN. (2) The experimental results for all indicators of RCGAN surpass those of RGGNN, indicating that the contribution of GGNN to the model’s performance is overall greater than that of CGAN. One potential reason is that entity-level context dependence can encapsulate basic graph structure  and fine-grained semantic information \cite{INFER}.  (3) The performance of CADRL surpasses that of the variant models, underscoring the crucial role of the mutual effect between two-level contextual dependencies in enhancing recommendation accuracy.

\begin{figure*}
	\centering
	\subfigure[NDCG]
	{
		\centering
		\includegraphics[width=0.231\linewidth,height=2.7cm]{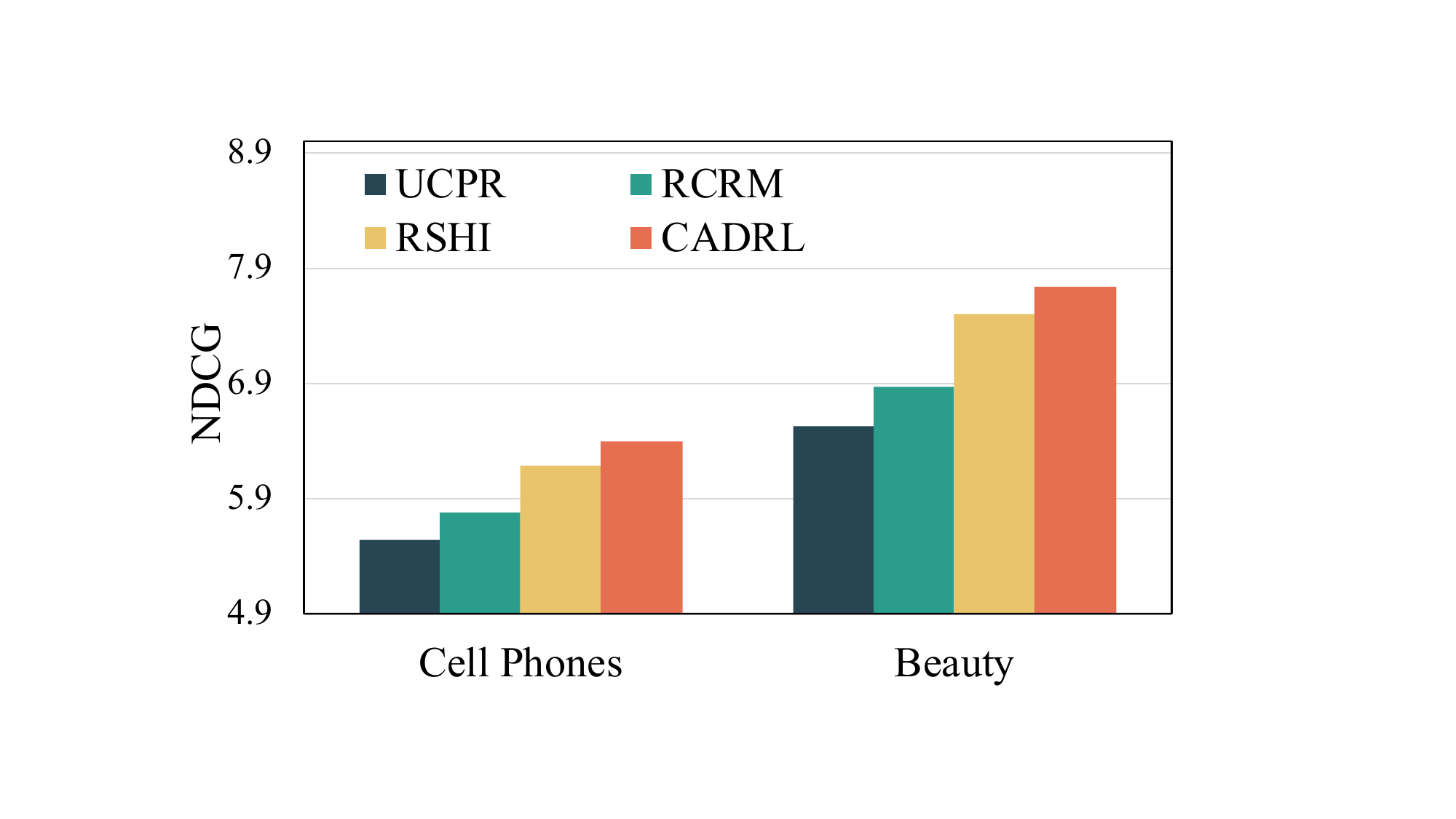}
	}
	\subfigure[Recall]
	{
		\centering
		\includegraphics[width=0.231\linewidth,height=2.7cm]{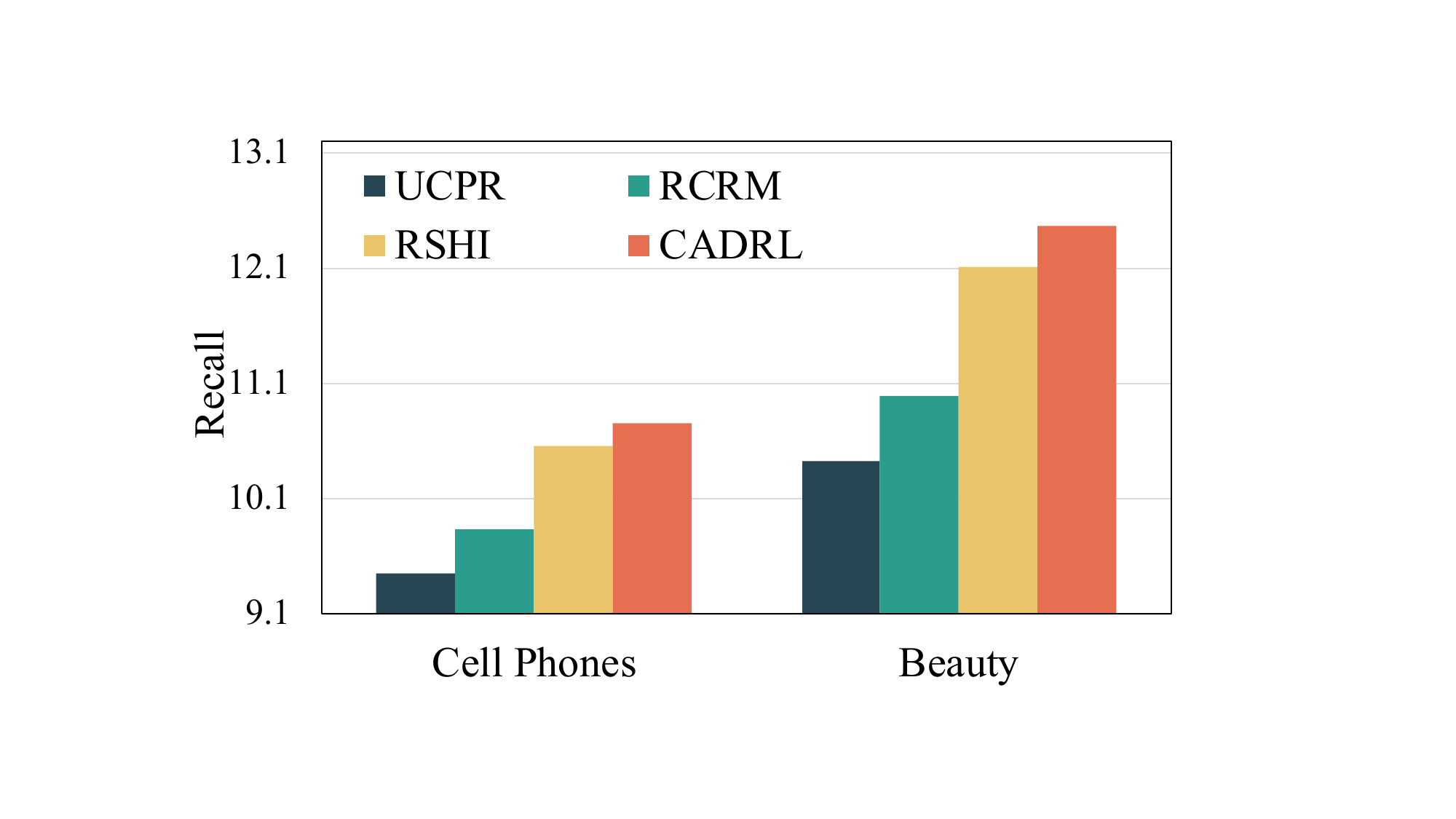}
	}
 \subfigure[HR]
	{
		\centering
		\includegraphics[width=0.231\linewidth,height=2.7cm]{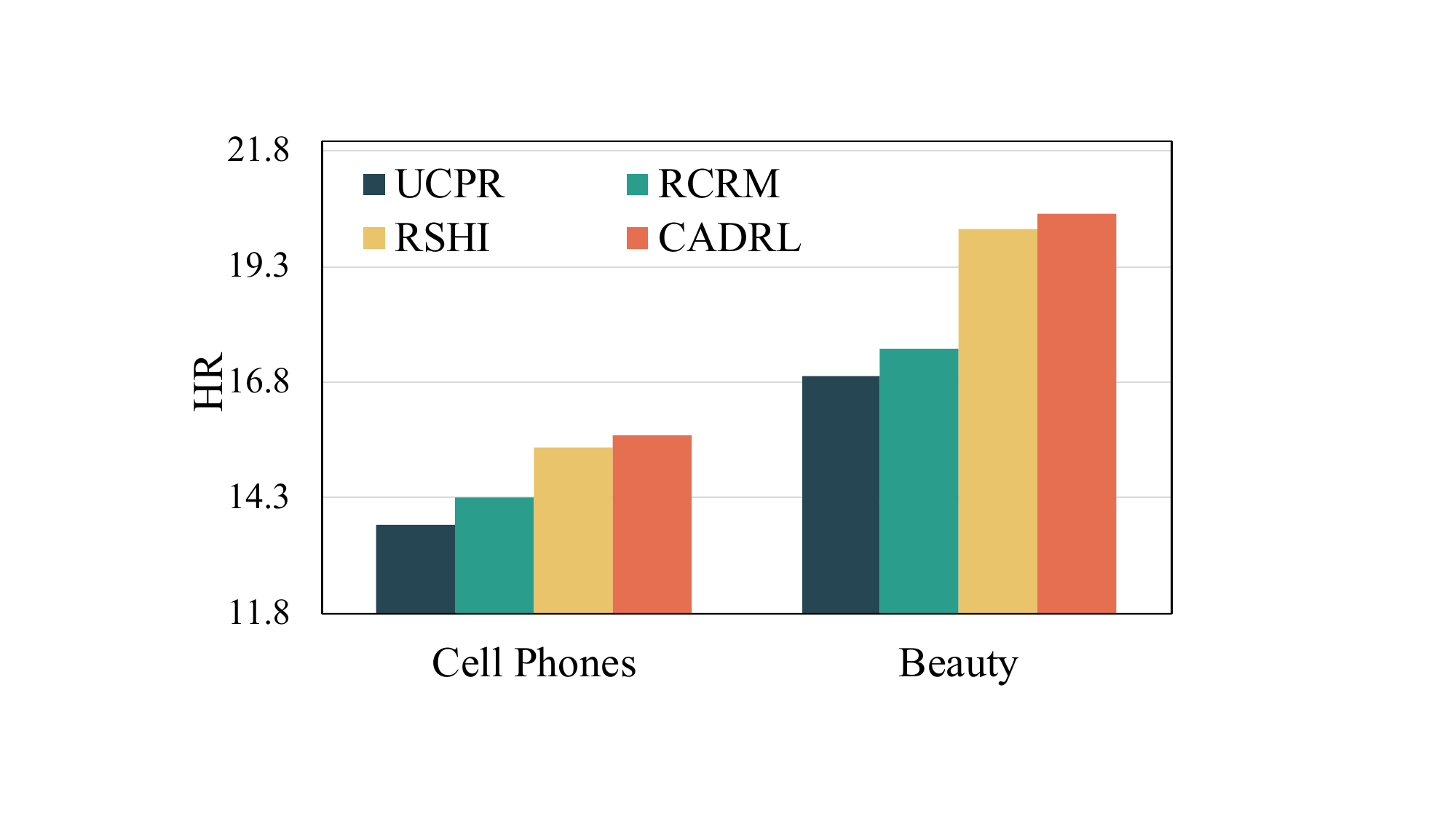}
	}
 \subfigure[Precision]
	{
		\centering
		\includegraphics[width=0.231\linewidth,height=2.7cm]{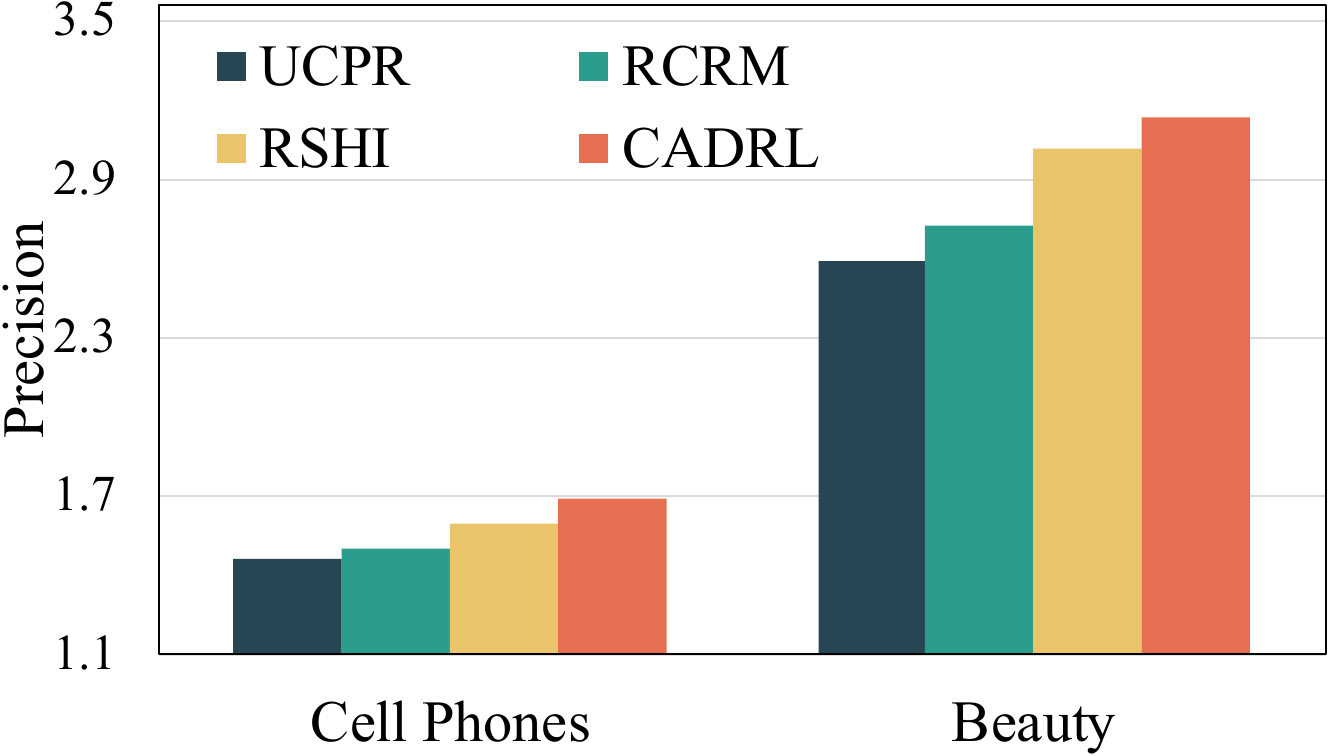}
	}
	\caption{Effectiveness analysis of different modules in DARL on different datasets.}
 
	\label{fig4}
\end{figure*}

\begin{figure*}
	\centering
	\subfigure[Clothing]
	{
		\centering
		\includegraphics[width=0.32\linewidth]{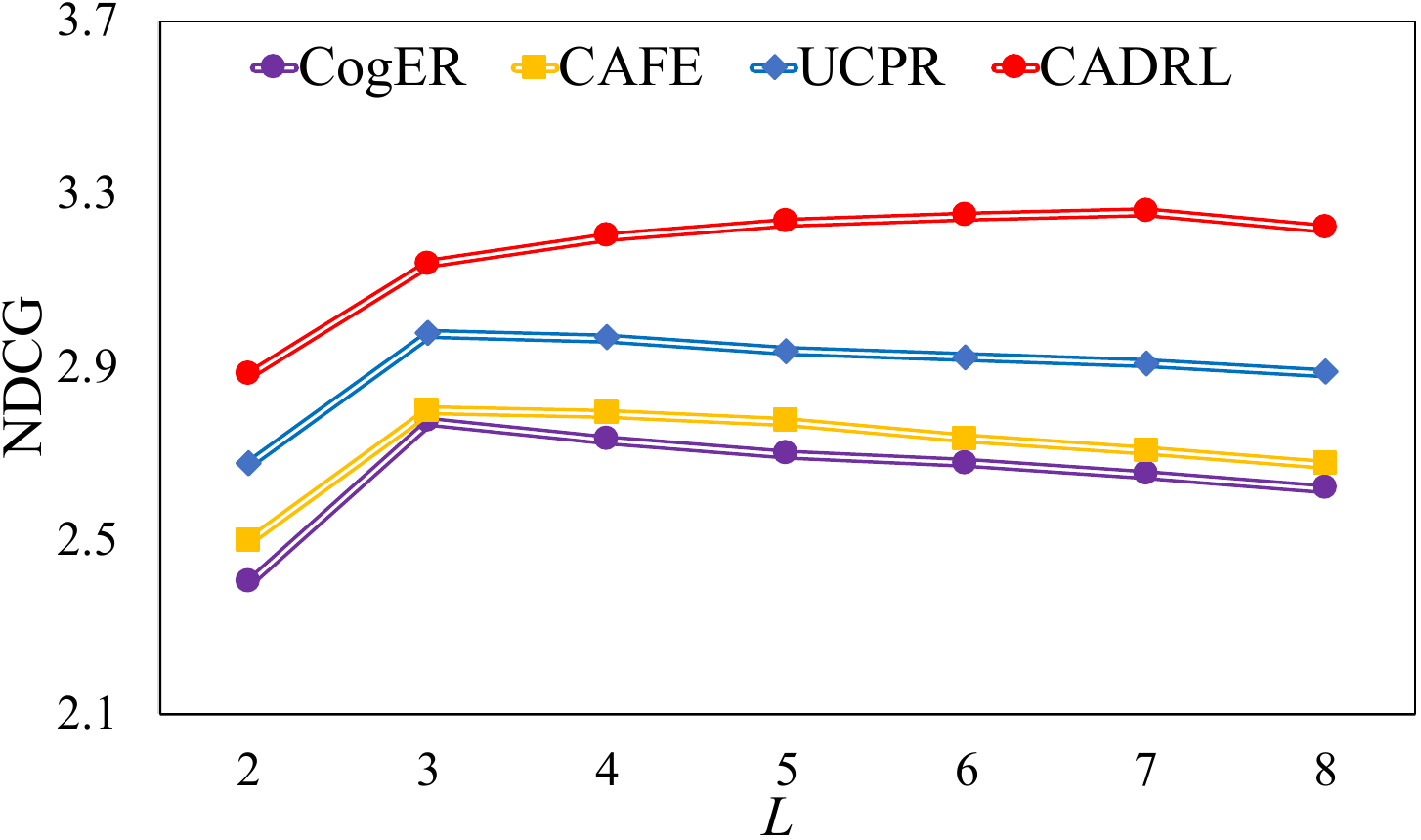}
	}
	\subfigure[Cell Phones]
	{
		\centering
		\includegraphics[width=0.31\linewidth]{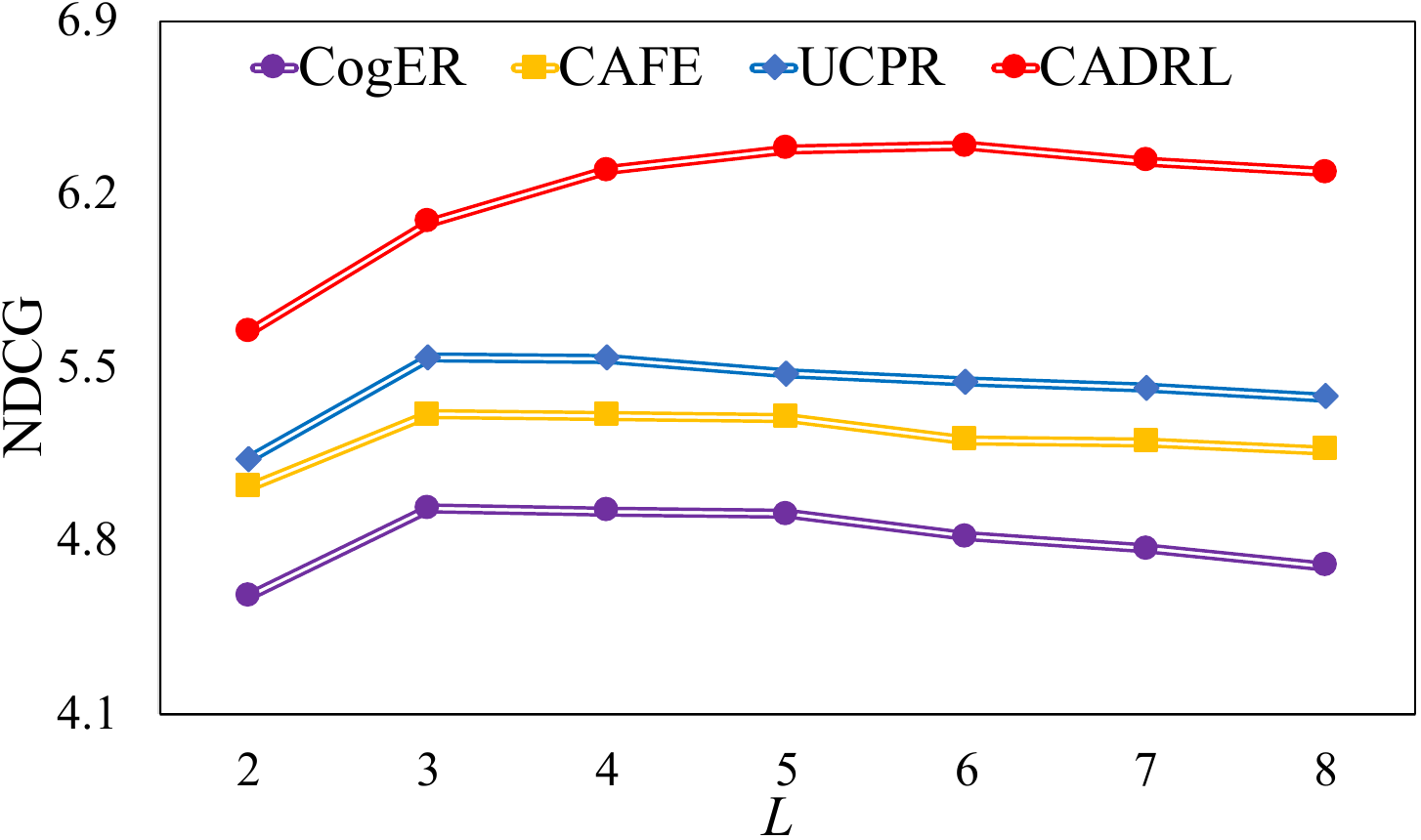}
	}
	\subfigure[Beauty]
	{
		\centering
		\includegraphics[width=0.31\linewidth]{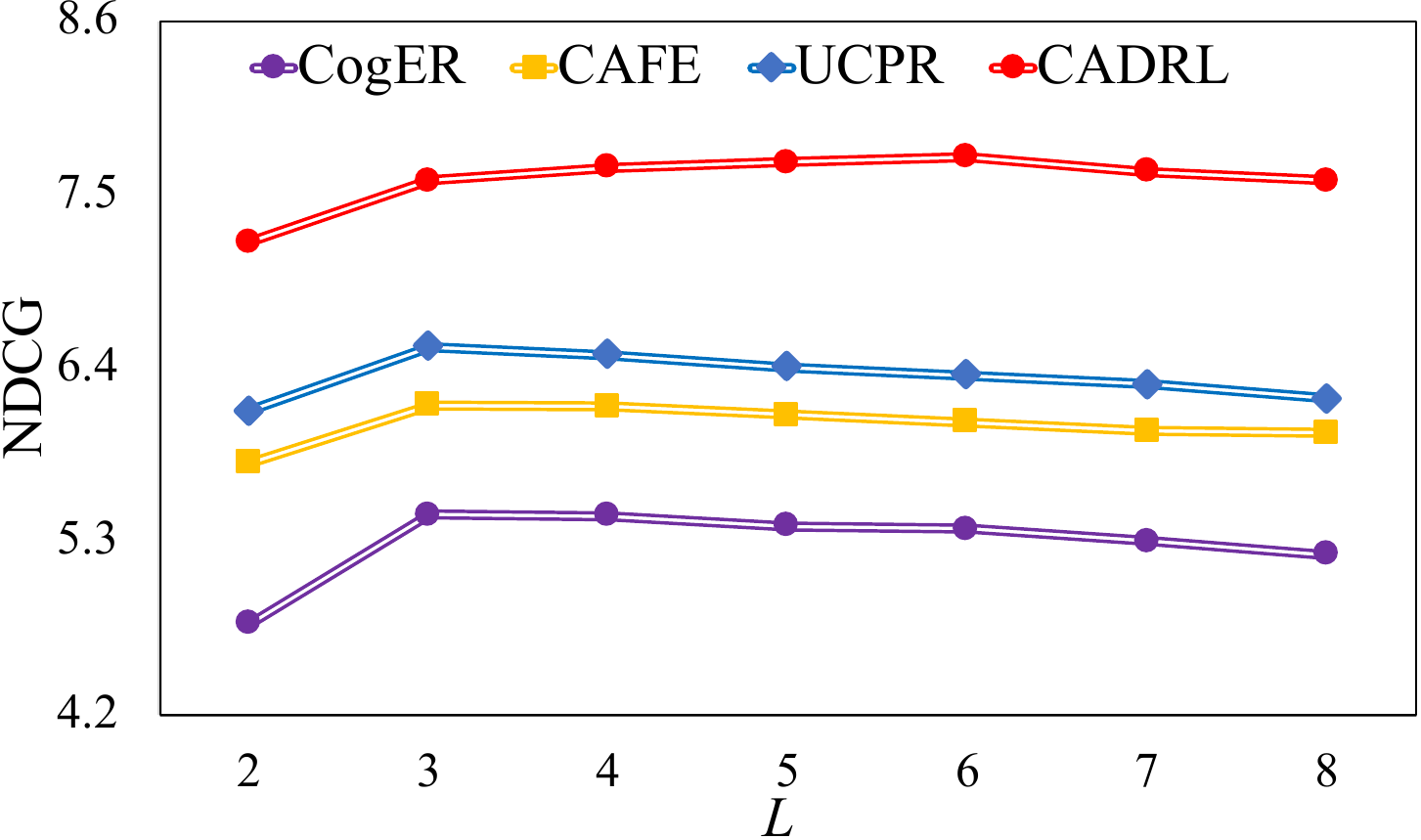}
	} 
	
	\caption{NDCG of the varying  recommendation step $L$ for RL-based models.}
	\label{fig5}
 \vspace{-0.2cm}
\end{figure*}

\subsubsection{The contribution of various modules in DARL}

The innovation of DARL is that the two agents collaboratively utilize historical interaction information and  partner rewards. Considering the necessity of reward functions for the RL framework, we design the following variant models to further study SPN and CRM by removing the technical modules.  (1) RSHI: the shared historical information of the dual agents is entirely removed, i.e., \textbf{y} is not calculated in Eq. (15) and (16). By comparing the performance of RSHI and CADRL, we can analyze the impact of the collaboration strategy in the SPN module on CADRL.  (2) RCRM: we remove the partner rewards in all reward functions of this variant, i.e., only the terminal reward is included in Eq. (20) and (21).

As illustrated in Fig. \ref{fig4}, we have the following insightful analysis: (1) Undoubtedly, CADRL maintains the highest performance among all methods, demonstrating its recommendation robustness. (2) The performance of RSHI is generally higher than that of RCRM but lower than that of CADRL. The reason is that the absence of shared historical information hinders DARL from capturing global preference sequences, thereby diminishing recommendation accuracy  \cite{PGPR}. (3) 
 RCRM outperforms the SOTA baseline UCPR, but its reliance on binary terminal rewards leads to the sparse reward dilemma. In addition, compared with RCRM, the performance advantage of CADRL illustrates that the collaborative reward mechanism in the DARL component has the ability to solve this dilemma.

\subsection{Path Length Study (RQ5)}
Existing RL-based explainable recommendation methods follow the idea of multi-hop reasoning over KGs \cite{KGzsf} \cite{multihopKGR} to limit  the recommendation path length. We argue that explainable recommendation methods over KGs encounter a more complex real-world scenario than multi-hop reasoning, where a user’s interaction sequence often exceeds three hops \cite{longseq}. To evaluate the rationality and the effectiveness of path length, we analyze the recommendation performance by extending the path length of various RL-based explainable recommendation methods in this subsection.

The experimental results are shown in Fig. \ref{fig5}, from which we derive the following insightful analysis. (1) CADRL breaks through the path-length limitation of existing RL-based explainable recommendation methods (i.e., their maximum path length is 3), achieving superior recommendation performance in longer paths. This success comes from the dual-agent RL collaboratively exploiting the latent semantic combination over extended paths to maintain performance advantages on all datasets. (2) The optimal path length of CADRL on the Clothing, Cell Phones and Beauty datasets are 7, 6 and 6, respectively. The growth trend accelerates initially and then decelerates, eventually turning negative upon reaching the optimal value. This phenomenon is attributed to the noise intervention along the extended path in the recommendation process.
(3) Existing RL-based baselines set the maximum path length to 3 through a hyperparameter. To compare the impact of path length, we increase the maximum length in the baselines to observe the experimental result. In Fig. 5, the baselines produce a large number of incorrect results and reduce their performance when the maximum length exceeds three hops. This is because these single-agent RL baselines suffer from the sparse reward dilemma and semantic dilution over extended paths \cite{RLsurvey1}. 

\begin{figure*}
	\centering
	\subfigure[Trade-off Factor $\delta$]
	{
		\centering
		\includegraphics[width=0.3\linewidth,height=3.4cm]{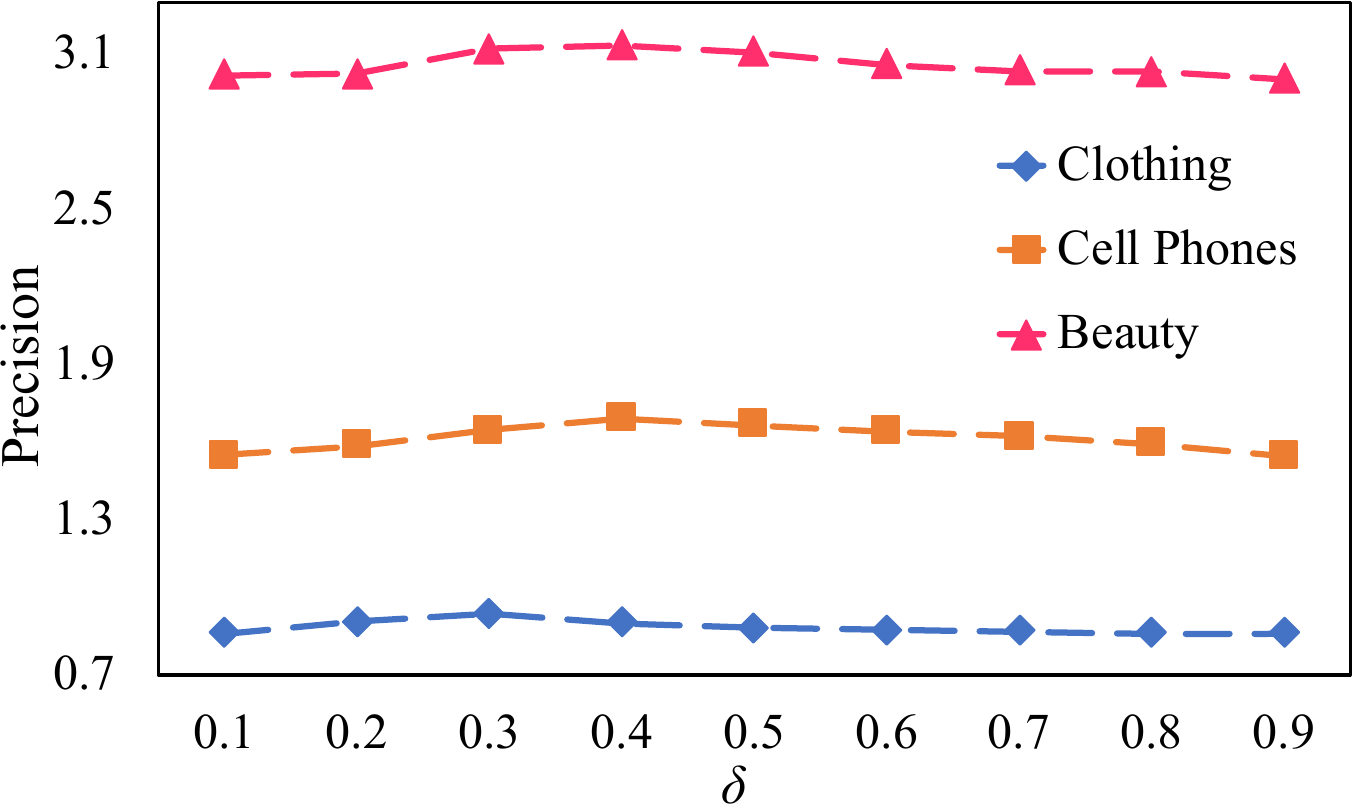}
	}
	\subfigure[Reward Discount Factor  $\alpha_{p_e}$]
	{
		\centering
		\includegraphics[width=0.3\linewidth,height=3.4cm]{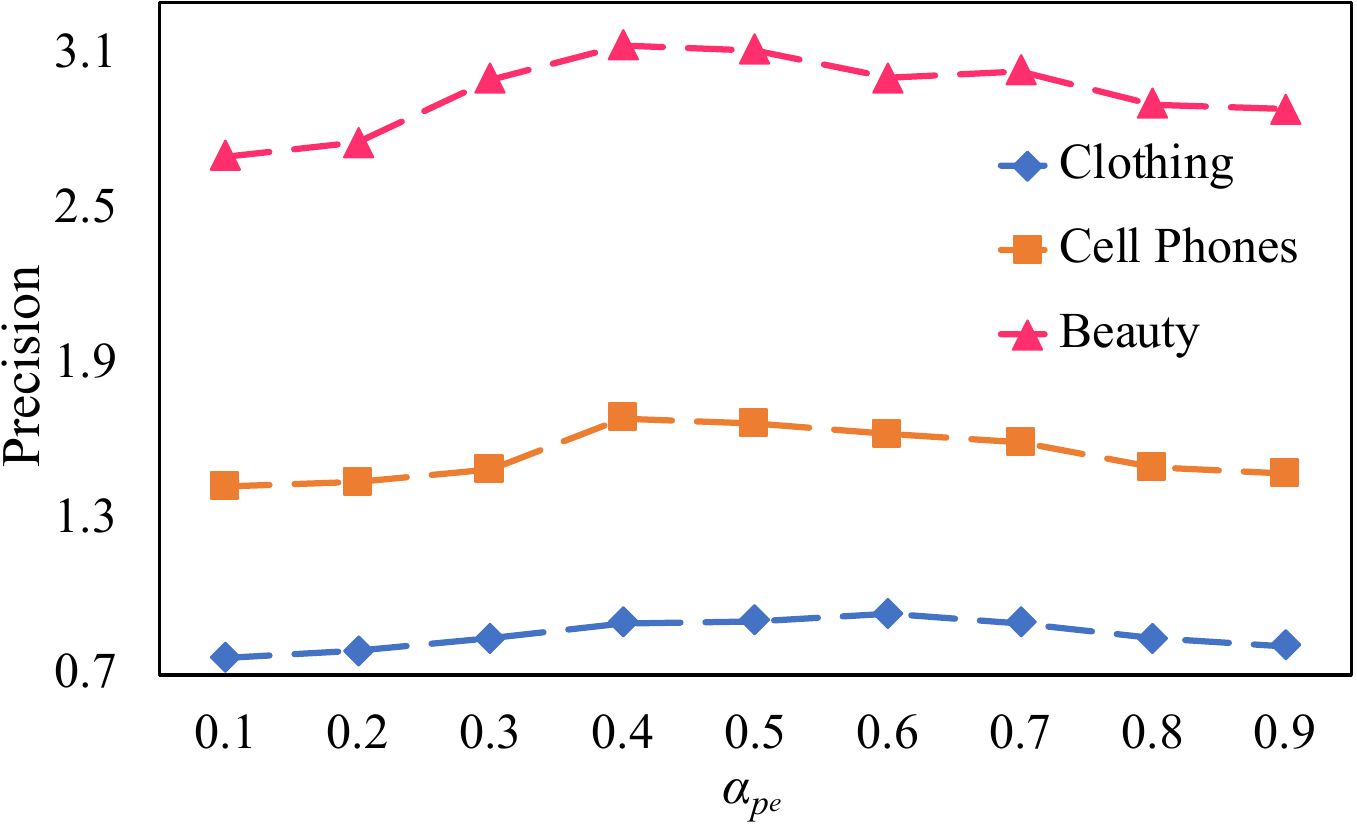}
	}
 \subfigure[Reward Discount Factor $\alpha_{p_c}$]
	{
		\centering
		\includegraphics[width=0.3\linewidth,height=3.4cm]{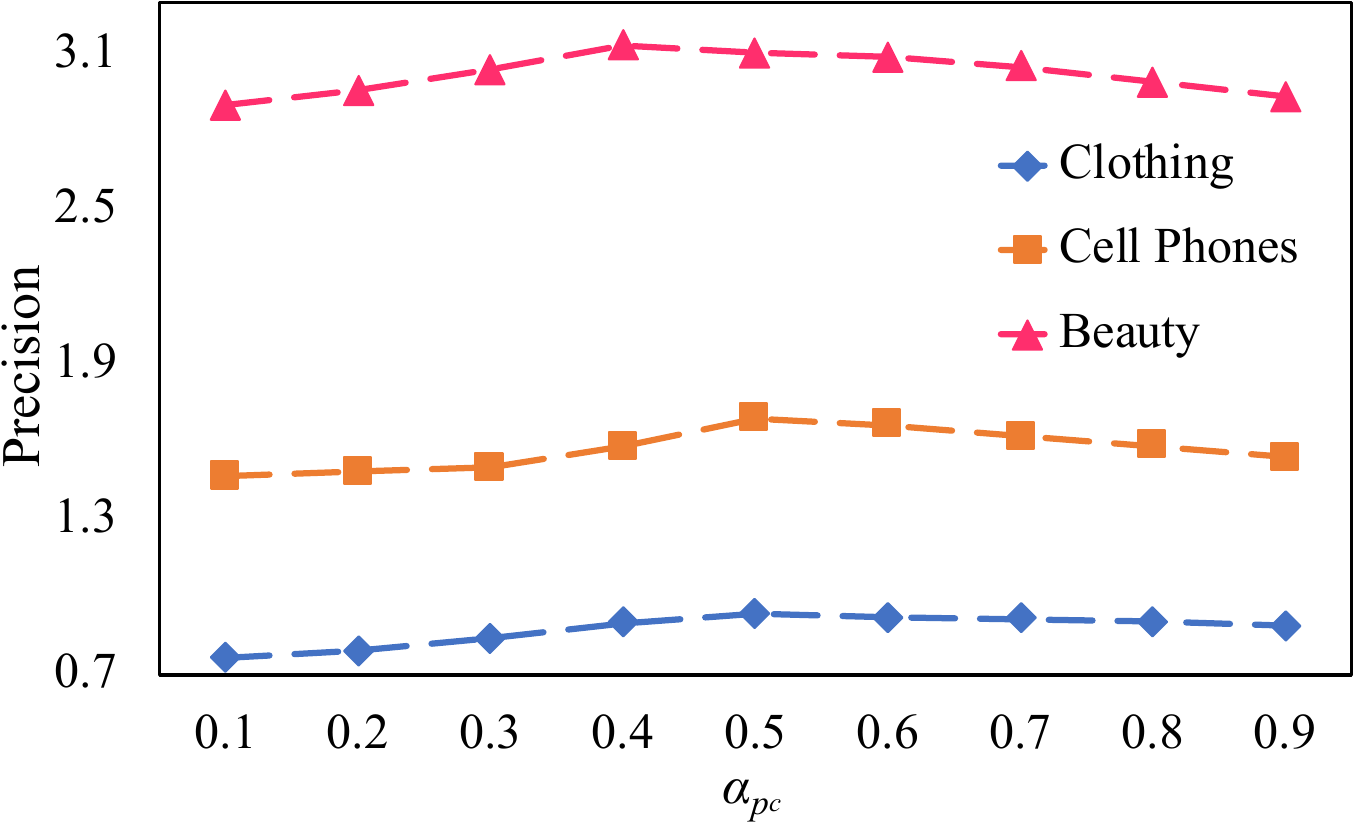}
	}
	\caption{The performance of CADRL  with the varying number of key hyper-parameters on different datasets.}
	\label{fig6}

\end{figure*}

\begin{figure*}
 \centering 

  \includegraphics[width=0.92\linewidth,height=6.2cm]{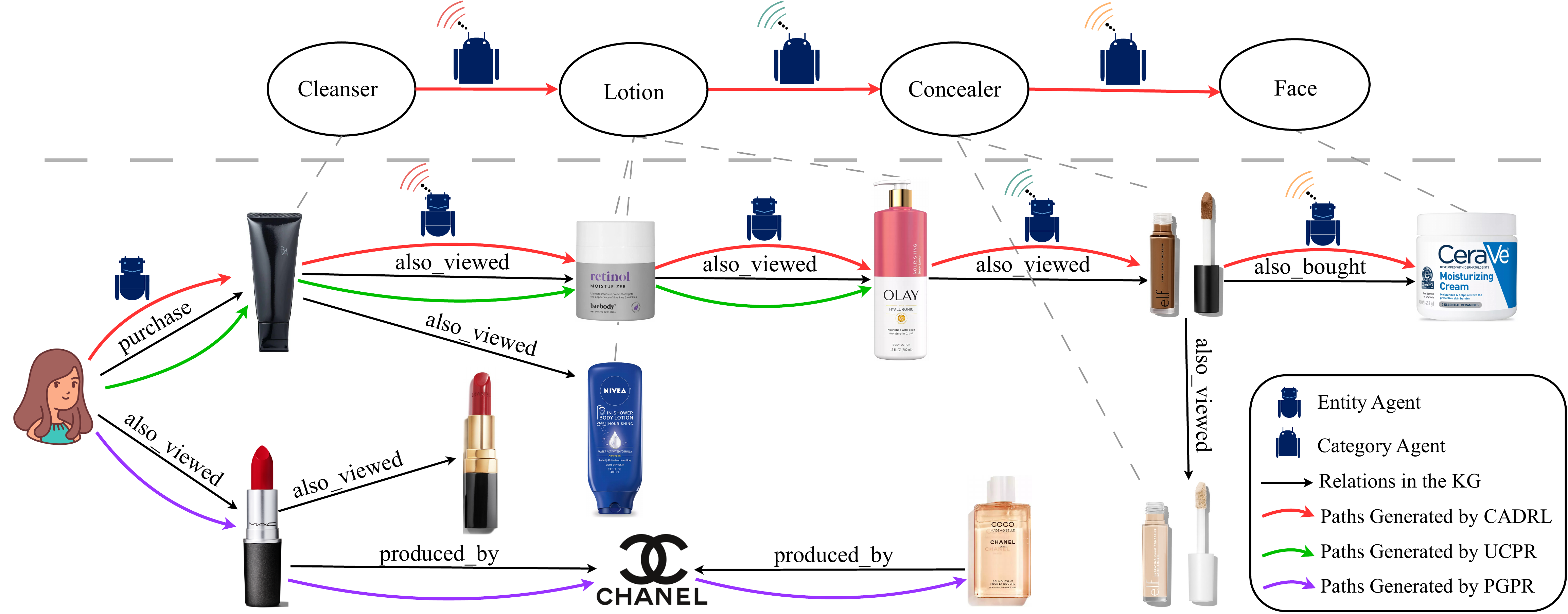}
 \hfill 
 \caption{Case study on Beauty.}
 \label{fig7}
 \vspace{-0.4cm}
\end{figure*}


 \subsection{Impact of Parameters (RQ6)} 
To answer RQ6, we perform parameter sensitivity analysis in this subsection. Notably, we investigate the performance  fluctuations of CADRL with the most critical hyper-parameters due to space constraints.
  \subsubsection{Impact of the trade-off factor $\delta$}
Fig. \ref{fig6}(a) illustrates the experimental results with varying trade-off factors $\delta$. As the value of $\delta$ changes, CADRL exhibits varying degrees of fluctuation in NDCG across different datasets. Specifically, the optimal $\delta$  values are 0.3, 0.4, and 0.4 for Clothing, Cell Phones, and Beauty, respectively. The optimal  $\delta$ value on Clothing is slightly lower than that on the other datasets. The above experimental results indicate that the number of categories is negatively correlated with the value of $\delta$. This is because a small $\delta$ value prevents CADRL from over-relying on category information when generating item representations in the dataset with more categories. 


 \subsubsection{Impact of the reward discount factor $\alpha_{p_e}$}
To further investigate the collaborative reward mechanism, we evaluate the impact of the reward discount factor $\alpha_{p_e}$ in Eq.(20). As illustrated by the results in Fig. \ref{fig6}(b), the optimal $\alpha_{p_e}$ values are 0.6, 0.4, and 0.4 for Clothing, Cell Phones, and Beauty, respectively. The key observations are as follows. (1) The $\alpha_{p_e}$ value is the highest on the Clothing dataset. One potential reason is that an appropriate $\alpha_{p_e}$ value can effectively alleviate the negative impact of item sparsity within each category on the path consistency of dual agents in this dataset. (2) The performance of CADRL fluctuates obviously on the Beauty dataset due to the data sparsity.  Nevertheless, the performance of CADRL across different $\alpha_{p_e}$ values generally surpasses that of existing SOTA baselines, which further demonstrates the robustness of CADRL performance.

\subsubsection{Impact of the reward discount factor $\alpha_{p_c}$}

Similarly, we investigate the optimal $\alpha_{p_c}$ value on different datasets.  The experimental results displayed in Fig. \ref{fig6}(c) reveal that CADRL’s performance initially improves and then declines across all datasets. This suggests that an appropriate $\alpha_{p_c}$ value can enhance the effectiveness of stage-wise guidance from the category agent. Consequently, the optimal values are determined to be 0.5, 0.5, and 0.4 for Clothing, Cell Phones, and Beauty, respectively.

\subsection{Case Study (RQ7)}

To investigate how CADRL utilizes the entity and category levels to perform explainable recommendations on KGs, we provide a real-world case study as shown in Fig. \ref{fig7}. The area above the grey horizontal line represents the abstract categories to which the items belong, while the area below is a KG fragment in the Beauty dataset. We derive the following key insights by observing these intuitive recommendation paths.

(1) Compared with PGPR and UCPR using the optimal path length setting, CADRL can use a 5-hop path to accurately find a suitable item for the user. The category agent captures the cross-selling opportunities between Cleanser and Face, while the entity agent combines the semantic associations of items and purchase behaviors in KG by receiving guidance and sharing path information. Dual agents collaboratively explore the next step at different levels until they reach the final goal, naturally forming an explainable provenance for the recommendation results. (2) The green and purple paths in Fig. 7 intuitively demonstrate the ``myopic" limitation of existing RL-based explainable recommendation methods. This can explain why there are many meaningless paths and false results in the output of existing RL-based methods over KGs. In contrast, the category agent can act as a ``myopic glasses" for the entity agent in our proposed model, further correcting CADRL's ``visual acuity"   (i.e., accurately find potential recommendation results beyond 3 hops).

\section{Conclusion And Future Work}
In this work, we study the problem how to leverage contextual dependencies to  efficiently conduct explainable recommendations without relying solely on short paths. To address this issue, we propose CADRL, which comprises CGGNN and DARL. Specifically, CGGNN captures fine-grained contextual dependencies within the neighboring structure from both the entity and category levels, generating high-order representations for items. To provide users with more suitable items by diverse semantic combinations on KGs, DARL introduces a dual-agent RL structure with a collaborative mechanism to explore long paths from the category and entity levels. Through extensive experiments on several real-world datasets, we draw two valuable conclusions. First, CADRL surpasses SOTA baselines in terms of efficiency and effectiveness in explainable recommendation tasks. Second, CADRL can infer more suitable items using long paths with the support of a collaborative dual-agent structure. In future work, we would like to employ large language models to perform personalized recommendations by modeling the complex evolution of user interests over KGs. 

\newpage

\bibliographystyle{IEEEtranS}

\bibliography{conference_101719}

\end{document}